\documentclass[aps,prd,notitlepage,nofootinbib,superscriptaddress,twocolumn,floatfix]{revtex4-2}

\pdfoutput=1

\usepackage{graphicx}
\usepackage{soul}
\usepackage{subfigure}
\usepackage{latexsym}
\usepackage{amssymb}
\usepackage{amsmath}
\usepackage{color}
\usepackage{array}
\usepackage{mathrsfs}
\usepackage{xparse}
\usepackage{bbding}
\usepackage{enumitem}
\usepackage{pifont}
\usepackage{enumerate} 
\usepackage{color}
\usepackage{mathrsfs}
\usepackage{xcolor}
\usepackage{changes}
\usepackage[colorlinks=true,linkcolor=blue,urlcolor=blue,citecolor=blue]{hyperref}
\usepackage[a4paper,margin=1in]{geometry}
\usepackage{amsmath,amssymb,mathtools}
\usepackage{bm}
\usepackage{enumitem}
\usepackage{hyperref}
\usepackage{physics}
\usepackage{orcidlink}

\begin{document}

 \newcommand{\bq}{\begin{equation}}
 \newcommand{\eq}{\end{equation}}
 \newcommand{\bqn}{\begin{eqnarray}}
 \newcommand{\eqn}{\end{eqnarray}}
 \newcommand{\nb}{\nonumber}
 \newcommand{\lb}{\label}
 
\newcommand{\La}{\Lambda}
\newcommand{\be}{\nopagebreak[3]\begin{equation}}
\newcommand{\ee}{\end{equation}}

\newcommand{\ba}{\nopagebreak[3]\begin{eqnarray}}
\newcommand{\ea}{\end{eqnarray}}

\newcommand{\la}{\label}
\newcommand{\n}{\nonumber}
\newcommand{\su}{\mathfrak{su}}
\newcommand{\SU}{\mathrm{SU}}
\newcommand{\U}{\mathrm{U}}

\def\be{\nopagebreak[3]\begin{equation}}
\def\ee{\end{equation}}
\def\ba{\nopagebreak[3]\begin{eqnarray}}
\def\ea{\end{eqnarray}}
\newcommand{\f}{\frac}
\def\rmd{\rm d}
\def\lp{\ell_{\rm Pl}}
\def\d{{\rm d}}
\def\fe{\mathring{e}^{\,i}_a}
\def\fw{\mathring{\omega}^{\,a}_i}
\def\fq{\mathring{q}_{ab}}
\def\t{\tilde}

\title{Probing the Distribution and Nature of Dark Matter Around Supermassive Black Holes from EMRI and IMRI Gravitational Waves}

\author{Jared Fier}
\email{Jared_Fier@baylor.edu}
\affiliation{GCAP-CASPER, Department of Physics, Baylor University,
One Bear Place $\#$97316, Waco, Texas, 76798-7316, USA}

\author{Farah Abdelshahed}
\email{Farah_Abdelshahed1@baylor.edu}
\affiliation{GCAP-CASPER, Department of Physics, Baylor University,
One Bear Place $\#$97316, Waco, Texas, 76798-7316, USA}

\author{Lilly Kowalczyk}
\email{Lilly_Kowalczyk1@baylor.edu}
\affiliation{GCAP-CASPER, Department of Physics, Baylor University,
One Bear Place $\#$97316, Waco, Texas, 76798-7316, USA}

\author{Zhengfei Lyu}
\email{Zhengfei_Lyu1@baylor.edu}
\affiliation{GCAP-CASPER, Department of Physics, Baylor University,
One Bear Place $\#$97316, Waco, Texas, 76798-7316, USA}


\author{Anzhong Wang \orcidlink{0000-0002-8852-9966}}
\email{anzhong_wang@baylor.edu; the corresponding author}
\affiliation{GCAP-CASPER, Department of Physics, Baylor University,
One Bear Place $\#$97316, Waco, Texas, 76798-7316, USA}

\date{(\today)}

\begin{abstract}

The distribution of dark matter in the immediate vicinity of
supermassive black holes remains poorly understood, despite its
importance for both galaxy evolution and precision tests of gravity.
Future space-based gravitational-wave observatories offer a unique
opportunity to probe this relativistic regime through the inspiral of
compact objects into supermassive black holes. Building upon our
previously constructed exact Einstein-cloud solutions within General
Relativity, we develop a fully relativistic framework for
investigating the gravitational-wave signatures of collisionless
dark-matter halos surrounding supermassive black holes. The framework
provides a unified treatment of the orbital dynamics, adiabatic
inspiral, accumulated gravitational-wave cycles, waveform phase
evolution, accumulated signal-to-noise ratio, and waveform mismatch
for extreme- and intermediate-mass-ratio inspirals (EMRIs/IMRIs). As
a representative application, we specialize the formalism to Model~I
and investigate its observational consequences. We show that
relativistic dark-matter halos can produce measurable modifications
to all four gravitational-wave diagnostics considered in this work,
namely the accumulated gravitational-wave cycles, waveform phase,
signal-to-noise ratio, and waveform mismatch, all of which lead to
consistent conclusions regarding detectability. By separating the
conservative modification of the spacetime geometry from the
dissipative effect of relativistic dynamical friction, we find that
the observable signatures are overwhelmingly dominated by the former,
whereas the latter remains negligible for the halo models considered
here. These results demonstrate that future space-based
gravitational-wave observations, such as those by
LISA, provide a promising means of probing the
relativistic distribution of dark matter around
supermassive black holes and highlight the potential of gravitational waves
as a powerful new probe of the distribution and
physical nature of dark matter surrounding
supermassive black holes.

\end{abstract}

\maketitle

\section{Introduction}

Determining the distribution and physical nature of dark matter (DM)
in the relativistic environments surrounding supermassive black holes
(SMBHs) is one of the outstanding challenges in modern astrophysics
and cosmology. Although compelling observational evidence accumulated
over the past several decades has established that dark matter
constitutes approximately $85\%$ of the matter content of the
Universe, its microscopic nature remains unknown and its distribution
in the strong-gravity regime surrounding SMBHs remains poorly
constrained~\cite{Planck18,Bertone05,Feng10}. Since SMBHs reside at
the centers of most massive galaxies~\cite{KormendyHo13}, where dark
matter, stars, gas, and compact objects coexist and interact,
understanding the relativistic distribution of dark matter in these
environments is essential not only for revealing the properties of
dark matter itself, but also for understanding galaxy formation, the
growth of SMBHs, and the dynamical evolution of galactic nuclei.

Despite remarkable progress in observational astronomy, the
distribution of dark matter in the immediate vicinity of SMBHs remains
one of the least explored aspects of galactic astrophysics. Existing
observational constraints are primarily derived from galactic rotation
curves, gravitational lensing, the cosmic microwave background, and
large-scale structure~\cite{Rubin80,Clowe06,Planck18}, which probe
dark matter on galactic or cosmological scales. More recently,
precision measurements of stellar orbits around Sagittarius
A$^\ast$~\cite{GRAVITY18,GRAVITY20} and horizon-scale images obtained
by the Event Horizon Telescope Collaboration~\cite{EHTM87,EHTSgrA}
have begun to provide unprecedented information about the environments
surrounding nearby SMBHs. These observations have also stimulated extensive investigations of
the relativistic distribution of dark matter around SMBHs and its
influence on stellar dynamics, black-hole shadows, quasinormal modes,
and other strong-field observables
~\cite{SFW13,SABB22,Cardoso22,ShenA,ShenB,Maeda}.
Although these observations have
significantly improved our understanding of the environments
surrounding nearby SMBHs, they probe different physical processes and
therefore provide only indirect constraints on the relativistic
distribution and physical properties of dark matter. Consequently, the
existence, distribution, and physical nature of dark matter around
SMBHs remain among the least understood aspects of galactic
astrophysics.

Complementary to electromagnetic observations, gravitational-wave (GW)
astronomy offers a fundamentally new opportunity to probe the
relativistic environments surrounding SMBHs. In particular,
extreme- and intermediate-mass-ratio inspirals (EMRIs/IMRIs),
consisting of a stellar-mass compact object orbiting a SMBH, are among
the most important sources for future space-based GW detectors such as
the Laser Interferometer Space Antenna (LISA). Unlike comparable-mass
compact binaries, EMRIs and IMRIs remain in the sensitive frequency
band for months to years, allowing even extremely small environmental
perturbations to accumulate coherently in the emitted waveforms.
During the inspiral, the compact object completes millions of orbital
cycles before plunging into the central SMBH, producing long-duration
GW signals whose phases can be measured with extraordinary precision.
Consequently, environmental effects that are negligible for other GW
sources may produce measurable modifications to the accumulated phase,
waveform morphology, and signal-to-noise ratio, making EMRIs and
IMRIs unique laboratories for probing the strong-field environments
surrounding SMBHs. This remarkable sensitivity has stimulated
extensive investigations of environmental effects on GW signals,
including accretion disks, dark matter halos, magnetic fields, and
other astrophysical perturbations~\cite{Babak17,Berry23}.

Motivated by the prospect of probing dark matter through strong-field
observations, considerable effort has been devoted to developing
physically motivated relativistic models of collisionless dark matter
around SMBHs within General Relativity. 
 A pioneering study by Gondolo and
Silk~\cite{GS99} investigated the adiabatic growth of SMBHs inside
dark-matter halos and predicted the formation of dense dark-matter
spikes surrounding the central black holes. Their analysis, based
primarily on Newtonian gravity with relativistic corrections, showed
that the dark-matter density must vanish inside $r=8M$, where stable
bound orbits no longer exist. Subsequently, Sadeghian
\textit{et al.}~\cite{SFW13} carried out a fully
general relativistic treatment of collisionless particle dynamics in
Schwarzschild spacetime and demonstrated that the physically relevant
inner boundary is located at $r=4M$. More recently, Speeney
\textit{et al.}~\cite{SABB22} further investigated the relativistic
phase-space distribution of collisionless dark matter around SMBHs and
confirmed that the density must vanish for $r<4M$. These studies
established an important physical picture: the distribution of
collisionless dark matter in the strong-gravity regime is governed by
the existence of stable bound orbits and therefore cannot, in
general, be represented by phenomenological halo profiles obtained by
simply extrapolating galactic-scale density distributions into the
relativistic regime.

Motivated by these developments, we recently constructed a family of
exact analytical solutions describing spherically symmetric
collisionless dark-matter halos surrounding Schwarzschild black holes
within General Relativity~\cite{ShenA}. The corresponding Einstein
clouds satisfy the relativistic inner boundary condition
$$
\rho(r< 4M)=0,
$$
and provide a unified description of both cored
and cuspy dark-matter distributions through a common analytical
parameterization. Owing to their exact analytical nature, these
solutions provide a natural relativistic framework for quantitatively
investigating a broad range of strong-field astrophysical phenomena
associated with dark matter surrounding SMBHs.

While Ref.~\cite{ShenA} established a relativistic framework for
constructing physically motivated Einstein-cloud solutions, the
ultimate goal is to determine whether future observations can distinguish 
different dark-matter distributions and thereby probe both the distribution and 
the physical nature of dark matter surrounding SMBHs. In this paper,
building upon the Einstein-cloud solutions developed in
Ref.~\cite{ShenA}, we develop a fully relativistic framework for
studying the influence of collisionless dark-matter halos on the
gravitational-wave emission from EMRIs/IMRIs. We derive the general equations
governing the orbital evolution, accumulated gravitational-wave
cycles, waveform phase, signal-to-noise ratio, and waveform mismatch,
and illustrate the general formalism by applying it to Model~I of the
Einstein-cloud solutions. By separating the conservative modification
of the background spacetime geometry from the dissipative
contribution of relativistic dynamical friction, we identify the
dominant physical mechanisms responsible for the observable
gravitational-wave signatures. Our results demonstrate that future
space-based gravitational-wave observations provide a promising means
of probing the distribution and physical nature of dark matter around
SMBHs. The framework developed here also provides a basis for future
multimessenger investigations of relativistic dark-matter halos using
gravitational waves together with complementary probes such as
black-hole shadows, stellar dynamics, and quasinormal modes.

The remainder of this paper is organized as follows. In Sec.~II we
briefly review the Einstein-cloud spacetime adopted in the present
work. Section~III develops the general relativistic framework for
describing the orbital evolution and gravitational-wave emission of
EMRIs and IMRIs embedded in Einstein-cloud environments. The
corresponding gravitational-wave signatures are presented in
Sec.~IV. Finally, Sec.~V summarizes our main conclusions and
discusses future directions. An appendix is also included, in which the expansions 
of the relevant physical quantities, including the metric functions, are given in 
terms of the small parameter $r/a$, which is of the order of $10^{-10}$ for the systems considered in this paper.

\section{Spherically Symmetric Einstein Clouds Around Massive Black Holes}

We consider a static, spherically symmetric spacetime described by \cite{ShenA,ShenB,Maeda}
\begin{equation}
ds^2=-f(r)dt^2+\frac{dr^2}{1-2m(r)/r}+r^2d\Omega^2,
\label{metric}
\end{equation}
where $f(r)$ is the lapse function, $m(r)$  the Misner--Sharp mass, and $d\Omega^2 \equiv \left(d\theta^2+\sin^2\theta\,d\phi^2\right)$.
In this paper, the matter surrounding the black hole is modeled as an Einstein cloud \cite{Cardoso22},
namely, a stationary ensemble of collisionless massive particles moving
on circular timelike geodesics. This description is physically motivated
by the standard cold-dark-matter picture, in which nongravitational
interactions and direct particle-particle encounters are sufficiently
rare that the collisional relaxation time is much longer than the
orbital and evolutionary timescales of the halo. The particles therefore
interact predominantly through the collective gravitational field and
may be treated within a collisionless, or Vlasov, description.
This approximation is expected to be particularly accurate for cold dark matter, 
whose nongravitational self-interaction cross section is observationally constrained to 
be sufficiently small that the halo evolves predominantly through collisionless gravitational dynamics.

At each radius, the orbital planes and the two possible directions of
motion are assumed to be distributed isotropically. Consequently, the
ensemble possesses no net angular momentum and remains static and
spherically symmetric, although the individual particles move on
circular geodesics. In the Einstein-cloud model considered here, the
particle four-velocity has no radial component, so there is no radial
momentum flux and the radial pressure vanishes identically. The orbital
motion, however, gives rise to a nonvanishing tangential velocity
dispersion and hence a nonzero tangential pressure. After averaging over
the particle ensemble, the coarse-grained stress-energy tensor takes the
form
\begin{equation}
T^\mu{}_\nu=
{\rm diag}\!\left[-\rho(r),\,0,\,P(r),\,P(r)\right],
\label{emt}
\end{equation}
where $\rho(r)$ and $P(r)$ denote the energy density and tangential pressure,
respectively, while the radial pressure vanishes identically.

The anisotropy of the stress-energy tensor in Eq.~(\ref{emt}) is therefore not
introduced through an ad hoc equation of state, but follows directly
from the microscopic dynamics of collisionless particles moving on
circular geodesics. Centrifugal support is provided entirely by
tangential orbital motion rather than by isotropic fluid pressure.
Consequently, the Einstein-cloud model provides a self-consistent
relativistic description of a collisionless dark-matter halo in which
the spacetime geometry and matter distribution are determined
simultaneously from Einstein's field equations.

Substituting Eqs.~(\ref{metric}) and (\ref{emt}) into Einstein's field
equations,
$G^\mu{}_\nu=8\pi T^\mu{}_\nu$ \footnote{In this paper we use the geometrized units, $G=c=1$, where $G$ and $c$ denote the Newtonian constant and speed of light, respectively.},
one finds that in the current case Einstein's field
equations have only   the following three
independent equations  \cite{ShenA,ShenB,Maeda}
\begin{align}
m'(r)&=4\pi r^2\rho(r), \label{eq:mprime}\\
\frac{f'(r)}{f(r)}
&=\frac{2m(r)}{r\,[r-2m(r)]}, \label{eq:fprime}\\
P(r)
&=\frac{r\rho(r)}{4}\frac{f'(r)}{f(r)}
=\frac{\rho(r)m(r)}
{2\,[r-2m(r)]}.
\label{eq:pressure}
\end{align}

Equations~(\ref{eq:mprime})--(\ref{eq:pressure}) provide three independent relations among the four unknown functions $f(r)$, $m(r)$, $\rho(r)$, 
and $P(r)$. Consequently, one additional physical input is required to close the system. Following our previous work~\cite{ShenA,ShenB}, we prescribe the dark-matter
density profile in the general form
\begin{equation}
\rho(r)=
\rho_0
\frac{\left(1-\dfrac{4M}{r}\right)^n H(r-4M)}
{\left(\dfrac{r}{a}\right)^\gamma
\left[1+\left(\dfrac{r}{a}\right)^\alpha\right]^{(\beta-\gamma)/\alpha}},
\label{eq:rho_general}
\end{equation}
where $H(x)$ denotes the Heaviside step function,
\begin{equation}
H(x)=
\left\{
\begin{array}{ll}
0, & x<0,\\[1mm]
1, & x\ge0.
\end{array}
\right.
\end{equation}
In Eq.~(\ref{eq:rho_general}), $\rho_0$ denotes the characteristic density of the halo,
$M$ the black hole mass, 
while $a$ is its characteristic scale radius. The three parameters
$(\alpha,\beta,\gamma)$ determine the overall shape of the density
distribution: $\gamma$ specifies the logarithmic slope of the inner
halo, $\beta$ determines the asymptotic outer slope, and $\alpha$
controls the sharpness of the transition between these two asymptotic
regions. The additional parameter
$n$ governs the behavior of the density near the relativistic inner
boundary $r=4M$, determining how smoothly the density approaches zero
there. Different choices of $(\alpha, \beta, \gamma, n)$ therefore
generate different relativistic dark-matter halo models.

Equation (\ref{eq:rho_general}) is a relativistic generalization of the widely used
double-power-law family of dark-matter density profiles \cite{Hernquist90,Zhao96}. By choosing
different values of the parameters $(\alpha,\beta,\gamma)$, it can
reproduce both cuspy and cored halo models commonly employed in
galactic dynamics \cite{BHS05,Freese09,SBD17,AM21}. In particular, when $\gamma = 0$, it corresponds to core models, while when $\gamma \not= 0$, it corresponds to cusp models.
This discrepancy, commonly known as the
core--cusp problem, remains one of the outstanding issues in modern
dark-matter astrophysics. Consequently, rather than restricting
ourselves to either cuspy or cored halos, we adopt the more general
family (\ref{eq:rho_general}), which encompasses both classes of models within a unified
framework while simultaneously incorporating the relativistically
motivated inner cutoff through the Heaviside factor.

It should be noted that collisionless cold-dark-matter
cosmological simulations generally predict cuspy inner density
profiles, such as the Navarro--Frenk--White (NFW) profile with
$(\alpha,\beta,\gamma) = (1, 3, 1)$ \cite{NFW97}, and the Hernquist model with $(\alpha,\beta,\gamma) = (1, 4, 1)$ \cite{Hernquist90}, whereas many observational studies of dwarf and
low-surface-brightness galaxies favor shallower, approximately
constant-density cores.

The Heaviside factor in Eq.~(\ref{eq:rho_general}) has a clear physical
origin. The pioneering work of Gondolo and Silk~\cite{GS99},
based on a Newtonian treatment supplemented by relativistic
corrections, investigated the adiabatic growth of SMBHs
inside collisionless dark-matter halos. Their analysis showed
that particle captured by the black hole naturally depletes the central
region of the halo and leads to an effective inner cutoff at
$r\simeq8M$.

Subsequently, Sadeghian et al. ~\cite{SFW13} developed a fully general relativistic description of collisionless dark matter around a Schwarzschild black hole by constructing the equilibrium phase-space distribution of bound particles in the Schwarzschild spacetime. The distribution function is defined only over the region of phase space corresponding to bound, non-captured geodesics. Consequently, after integrating over the allowed phase space, the resulting equilibrium density possesses no support for $r<4M$, so that the physically relevant inner boundary of a stationary collisionless halo is located at $r=4M$.  This fully relativistic result
was later confirmed independently by Speeney {\it et al.}~\cite{SABB22},
who also investigated its implications for realistic dark-matter halos
around SMBHs. 

Although the present Einstein-cloud model assumes that the constituent particles move on circular timelike geodesics, the condition 
\bq
\lb{eq2.8}
\rho\left(r < 4M\right)=0,
\eq
 is not introduced as an independent assumption. Rather, it is adopted from the fully relativistic equilibrium phase-space analyses of collisionless dark matter, in which the distribution function is supported only by bound, non-captured geodesics. Consequently, no stationary equilibrium density exists inside $r=4M$ \cite{SFW13,SABB22}. The Einstein-cloud model therefore incorporates this relativistic inner boundary explicitly through the Heaviside factor in Eq.~(\ref{eq:rho_general}).

Many dark-matter profiles widely used in galactic dynamics are designed
to describe halo structures on galactic scales and are not intended to
model the strong gravitational field in the immediate vicinity of a
black hole. Since these phenomenological profiles are not constructed from the
equilibrium phase-space distribution of collisionless particles in the
strong gravitational field of a black hole, they generally do not
incorporate the relativistic particle-capture boundary and therefore
predict a nonvanishing density arbitrarily close to the center when
extrapolated into the near-horizon region.

In our previous work~\cite{ShenA}, five exact Einstein-cloud solutions
were constructed from Eq.~(6). Among them, Model I possesses the
simplest analytical structure while retaining all the essential
physical characteristics of relativistic dark-matter halos, including
the physically motivated inner boundary at $r=4M$, and an exact analytical
 spacetime. 
Although the
gravitational-wave formalism developed in the following sections is
applicable to all Einstein-cloud solutions, when applying the formulas to specific spacetimes, in  this paper we
adopt Model I as a representative example \footnote{This model corresponds to the choice $\left(n, \alpha, \beta, \gamma\right) = (1, 1, 5, 1)$. So it represents a cusp model.}.
 It therefore serves as an ideal benchmark for illustrating how
relativistic dark-matter halos modify the orbital evolution and
gravitational-wave emission from EMRIs and IMRIs.

\section{Gravitational Waves from EMRIs/IMRIs and Observational Signatures}

In this section, we consider the gravitational waves emitted by a compact object, which is orbiting a SMBH surrounded by a dark matter halo. Since $ m_* \ll M$, as the leading-order approximation, we first ignore the backreaction of the object and consider it as a test particle  moving around the SMBH. Here $m_*$  denotes  the mass of the compact object. We assume that the  radius of its orbit is always greater than $4M$, so the compact object passes through the dark matter halo. Clearly, for such motions the gravitational drag is induced on the inspiraling object and changes its orbit and the gravitational waveforms emitted by it, depending on the nature and distribution of the dark matter halo. 

To make the formalism developed in this section as generally applicable
as possible, we do not restrict ourselves to the exact Einstein-cloud
solutions constructed in Ref.~\cite{ShenA}. Instead, we consider the
general case in which the dark matter halo is described by an Einstein
cloud. The resulting equations are therefore applicable to all
Einstein-cloud models, while the specific exact solutions presented in
Ref.~\cite{ShenA} will be used in Sec.~IV to illustrate their physical
implications.

 Before considering the dissipative evolution of the binary,
we first derive the properties of timelike circular geodesics
in the Einstein-cloud spacetime. These geodesics describe
the conservative orbital dynamics of the compact object in
the test-particle approximation and provide the foundation
for the adiabatic inspiral formalism developed in the
following subsections.

\subsection{Timelike Geodesics and Circular Orbits}
\label{sec:timelike_geodesics}

A test particle of rest mass $m_\star$ in the
static, spherically symmetric geometry described by Eq.(\ref{metric}) is given by the geodesic Lagrangian per unit rest mass $\mathcal{L}_{\rm geo} = \frac{1}{2} g_{\mu\nu}\dot{x}^{\mu} 
\dot{x}^{\mu}$, where $\dot{x}^{\mu} \equiv dx^{\mu}/d\tau$ with $\tau$ denoting the proper time of the test particle. 
However, because of spherical symmetry, the orbit can always be chosen to lie in the
equatorial plane, $\theta=\pi/2$, with $\dot\theta=0$.  Then, we find  
\begin{equation}
 2\mathcal{L}_{\rm geo}
 =-f(r)\dot t^{\,2}
 +\frac{\dot r^{\,2}}{1-2m(r)/r}
 +r^2\dot\phi^{\,2}.
\label{eq:geo_lagrangian}
\end{equation}
Clearly, the coordinates $t$ and $\phi$ are cyclic. As a result, their conjugate momenta are conserved, which gives the specific energy $\mathcal E$ and the
specific angular momentum $\mathcal L$,
\begin{equation}
 \mathcal E\equiv f(r)\dot t,
 \qquad
 \mathcal L\equiv r^2\dot\phi.
\label{eq:geo_constants}
\end{equation}
For a timelike geodesic the four-velocity satisfies
$u^\mu u_\mu=-1$. Substituting Eq.~\eqref{eq:geo_constants} into this
normalization condition yields the radial equation
\begin{equation}
 \dot r^{\,2}
 =\left(1-\frac{2m(r)}{r}\right) R(r;\mathcal E,\mathcal L),
\label{eq:radial_geodesic}
\end{equation}
where  
\begin{equation}
 \mathcal R(r;\mathcal E,\mathcal L)
 \equiv
 \frac{\mathcal E^2}{f(r)}-1-\frac{\mathcal L^2}{r^2}.
\label{eq:radial_potential}
\end{equation}

For a circular timelike geodesic,  $r=$ Constant, 
 both its radial velocity and radial acceleration must vanish, which are equivalently to
\begin{equation}
 \mathcal R(r)=0,
 \qquad
 \frac{\partial\mathcal R(r)}{\partial r}=0,
\label{eq:circular_conditions}
\end{equation}
at $r=$ Constant.  These two conditions can be cast explicitly in the form
\bqn
&& \frac{\mathcal E^2}{f}
 =1+\frac{\mathcal L^2}{r^2},
\label{eq:circular_condition_1}\\
&& \frac{\mathcal E^2 f'}{f^2}
 -\frac{2\mathcal L^2}{r^3}
 =0.
\label{eq:circular_condition_2}
\eqn
Solving Eqs.~\eqref{eq:circular_condition_1} and
\eqref{eq:circular_condition_2} gives the general circular-orbit formulas
\begin{equation}
 \mathcal E^2
 =\frac{2f^2}{2f-rf'},
 \qquad
 \mathcal L^2
 =\frac{r^3f'}{2f-rf'}.
\label{eq:EL_general_f}
\end{equation}
For the Einstein-cloud solutions considered in this paper, $f'(r)/f(r)$ is given by Eq.(\ref{eq:fprime}), which, together with Eq.~\eqref{eq:EL_general_f}, yields
\begin{align}
 \mathcal E^2
 &=f(r)\frac{r-2m(r)}{r-3m(r)},
\label{eq:circular_energy}\\
 \mathcal L^2
 &=\frac{r^2m(r)}{r-3m(r)}.
\label{eq:circular_angular_momentum}
\end{align}
These expressions require $r>3m(r)$ for a timelike circular orbit \cite{Maeda}.

Then, the coordinate angular velocity 
  is given by
\begin{equation}
 \Omega(r) \equiv\frac{d\phi}{dt}
 =\frac{\dot\phi}{\dot t}
 =\frac{f\mathcal L}{r^2\mathcal E}.
\label{eq:omega_definition}
\end{equation}
Substituting Eq.~\eqref{eq:EL_general_f} gives the well-known relation
$\Omega^2=f'/(2r)$. Equations~\eqref{eq:fprime},  ~\eqref{eq:circular_energy} and ~\eqref{eq:circular_angular_momentum} then yield
\begin{equation}
 \Omega^2(r)
 =\frac{f(r)m(r)}{r^2(r-2m(r))}.
\label{eq:circular_omega}
\end{equation}
Equation (\ref{eq:circular_omega}) generalizes Kepler's third law to the Einstein-cloud spacetime. All subsequent gravitational-wave observables depend on the halo through this modified orbital frequency.

Finally, the physical azimuthal speed measured by a static observer at the
same radius follows from the observer's proper time
$d\tau_{\rm stat}=\sqrt{f}\,dt$ and proper circumferential distance
$d\ell=r\,d\phi$
\begin{equation}
 v^2(r)
 \equiv\left(\frac{d\ell}{d\tau_{\rm stat}}\right)^2
 =\frac{r^2\Omega^2}{f(r)}
 =\frac{m(r)}{r-2m(r)}.
\label{eq:circular_velocity}
\end{equation}
Equations~\eqref{eq:circular_energy},
\eqref{eq:circular_angular_momentum},
\eqref{eq:circular_omega}, and
\eqref{eq:circular_velocity} are the basic quantities used below to construct
the adiabatic inspiral and gravitational waveform.

\subsection{Innermost Stable Circular Orbit}

The circular geodesics derived above exist only when
$r>3m(r)$. For gravitational-wave astronomy,
however, one is primarily interested in those circular
orbits that are dynamically stable under small radial
perturbations.
 In terms of the
radial function in Eq.~\eqref{eq:radial_potential}, stability requires the
circular orbit to lie on the stable side of the marginal point. 
The innermost stable circular orbit (ISCO) is the limiting
circular orbit at which radial stability is lost. It is therefore
determined by the three conditions
\begin{equation}
R=0,\qquad R'=0,\qquad R''=0.
\end{equation}

Equivalently, following Ref.~\cite{Maeda}, along the one-parameter
sequence of circular orbits, the ISCO is an extremum of
$L^2(r)$ (and also of $E^2(r)$), namely,
 ${d\mathcal L^2}/{dr}=0$.
From Eq.~\eqref{eq:circular_angular_momentum}, we obtain
\begin{equation}
 \frac{d}{dr}\left(\frac{r^2m}{r-3m}\right)
 =\frac{r\left(r^2m'+rm-6m^2\right)}
 {(r-3m)^2}.
\label{eq:dL2_dr}
\end{equation}
Consequently, the ISCO radius is the physically admissible root of 
\begin{equation}
 r^2m'(r)+rm(r)-6m^2(r)=0.
\label{eq:isco_condition}
\end{equation}
Among all mathematical roots of Eq. ({\ref{eq:isco_condition}), the physical
ISCO is the outermost root that lies outside the event
horizon and satisfies $r>3m(r)$.  In the Schwarzschild
limit, $m(r)=M$ and $m'(r)=0$, Eq.~\eqref{eq:isco_condition} immediately gives
$r_{\rm ISCO}=6M$, as expected.

\subsection{Inspiral}
\label{subsec:emri_inspiral}

An EMRI/IMRI consists of a compact object of mass
$m_\star$ moving around a massive black hole of mass $M$, with
$m_\star/M\ll1$.  On the orbital time scale the small body follows, to an
excellent approximation, a geodesic of the background spacetime.  On the
much longer radiation-reaction time scale, however, the orbit slowly loses
energy and angular momentum.  The motion may therefore be described as an
adiabatic sequence of nearly circular geodesics whose radius decreases with
time.  In the present problem there are two dissipative channels: the
emission of gravitational waves and the gravitational drag exerted by
the surrounding dark matter, usually referred to as dynamical friction
(DF). Throughout this work we assume that the inspiral remains
quasicircular, so that the orbital evolution is completely
characterized by the slowly varying orbital radius.

For a circular orbit the orbital energy is
\begin{equation}
 E_{\rm orb}(r)=m_\star\mathcal E(r),
 \label{eq:orbital_energy_emri}
\end{equation}
where $\mathcal E(r)$ is the specific energy in
Eq.~\eqref{eq:circular_energy}.  The adiabatic approximation is valid when
the inspiral time $r/|\dot r|$ is much longer than the orbital period
$2\pi/\Omega$.  Under this assumption, the secular evolution follows from
the energy-balance equation
\begin{equation}
 \frac{d{E}_{\rm orb}}{dt}
 =-\mathcal F_{\rm GW}-\mathcal F_{\rm DF},
 \label{eq:energy_balance_emri}
\end{equation}
where both luminosities are defined to be positive quantities.

\subsubsection{Gravitational Wave Luminosity}

The accelerated compact object generates spacetime perturbations that
propagate away as gravitational radiation.  The corresponding luminosity
$\mathcal F_{\rm GW}$ is the orbital energy carried by gravitational waves  per unit
coordinate time.  At leading, quadrupolar order, the luminosity of a
quasi-circular binary is \cite{Einstein18,Landau75,Maggiore08}
\begin{equation}
 \mathcal F_{\rm GW}
 =\frac{32}{5}m_\star^2 r^4\Omega^6.
 \label{eq:gw_flux_quadrupole}
\end{equation}
Equation
\eqref{eq:gw_flux_quadrupole} is used as the leading-order dissipative model,
while the conservative orbital quantities $\Omega(r)$ and
$\mathcal E(r)$ are evaluated in the exact halo geometry.  Substitution of
Eq.~\eqref{eq:circular_omega} gives
\begin{equation}
 \mathcal F_{\rm GW}
 =\frac{32m_\star^2}{5}
 \frac{f^3(r)m^3(r)}{r^2(r-2m(r))^3},
 \label{eq:gw_luminosity}
\end{equation}
which reduces to the familiar Schwarzschild quadrupole flux
$\mathcal F_{\rm GW}=(32/5)(m_\star/M)^2(M/r)^5$ when $m(r)=M$ and
$f(r)=1-2M/r$.  More accurate calculations may replace Eq.~\eqref{eq:gw_luminosity} 
by fluxes computed from black-hole perturbation
theory or gravitational self-force methods \cite{ZGW26} without
changing the general balance-law framework
developed below.
 
\subsubsection{Dynamical Friction Luminosity}

As the compact object moves through the dark-matter distribution, its
gravitational field focuses the surrounding particles and creates an
overdense wake behind it.  The gravitational attraction of this wake
opposes the motion of the compact object and removes orbital energy.  In a
collisionless medium, the standard Chandrasekhar drag force has magnitude
\cite{Chandrasekhar43}
\begin{equation}
 F_{\rm DF}^{\rm N}
 =\frac{4\pi m_\star^2\rho(r)\ln\Lambda}{v^2},
 \label{eq:chandrasekhar_force}
\end{equation}
where $\rho(r)$ is the local dark-matter density, $v$ is the physical orbital
speed measured by a static observer, and $\ln\Lambda$ is the Coulomb
logarithm.  The energy-loss rate measured with respect to the static observer's proper time
is $F_{\rm DF}^{\rm N}v$.  In this work we adopt the relativistic
Chandrasekhar-type prescription
\bqn
 \mathcal F_{\rm DF}
 &=&\frac{4\pi m_\star^2\rho(r)\ln\Lambda}{v}
 \sqrt{f(r)\left(1-v^2\right)}\,\xi(v), ~~~~
 \label{eq:df_flux_velocity_form}
 \eqn
 where   $\xi(v)$ is the relativistic correction factor
to Chandrasekhar's dynamical-friction formula \cite{SABB22}, given by
\bqn
 \xi(v)&=&\frac{(1+v^2)^2}{1-v^2}.
 \label{eq:df_flux_xiv_form}
\eqn
The redshift and Lorentz factors convert the locally evaluated drag into an
energy-loss rate with respect to the asymptotic time coordinate, while
$\xi(v)$ tends to unity in the nonrelativistic limit and enhances the drag
when the orbital velocity becomes relativistic.  Using
Eq.~\eqref{eq:circular_velocity}, one finds
\begin{equation}
 \frac{\sqrt{f(1-v^2)}}{v}
 =\sqrt{\frac{f(r-3m)}{m}}.
 \label{eq:df_velocity_relations}
\end{equation}
Consequently, the DF luminosity may be written as
\bqn
 \mathcal F_{\rm DF}
 &=&4\pi m_\star^2\rho(r)\ln\Lambda
 \sqrt{\frac{f(r)(r-3m(r))}{m(r)}}\nb\\
 && \times 
 \frac{(1+v^2(r))^2}{1-v^2(r)},
 \label{eq:df_luminosity}
\eqn
or, equivalently, in any algebraically reduced form obtained by inserting
$v^2=m(r)/(r-2m(r))$.  This contribution vanishes identically in the vacuum region
where $\rho=0$.  The use of a local Chandrasekhar-type prescription is an
approximation; it neglects the global response of the halo and possible
changes of the dark-matter distribution induced by the inspiral.  It is,
nevertheless, a useful first estimate of the environmental dissipation.

Since $E_{\rm orb}$ is a function of the instantaneous circular-orbit radius,
\begin{equation}
 \frac{dE_{\rm orb}}{dt}
 =\frac{dE_{\rm orb}}{dr}\,\dot r,
 \label{eq:orbital_energy_chain_rule}
\end{equation}
combining Eqs.~\eqref{eq:energy_balance_emri} and
\eqref{eq:orbital_energy_chain_rule}, we obtain the radial inspiral equation
\begin{equation}
 \dot r
 =-\frac{\mathcal F_{\rm GW}+\mathcal F_{\rm DF}}
 {dE_{\rm orb}/dr}.
 \label{eq:radial_inspiral}
\end{equation} 
For the stable
circular branch outside the ISCO, $dE_{\rm orb}/dr>0$, and therefore
$\dot r<0$, as required for an inward inspiral.  Eqs.~\eqref{eq:circular_omega} and ~\eqref{eq:radial_inspiral}, together with the
Einstein-cloud background determined in Sec. II,
completely specify the leading-order inspiral
trajectory and form the basis for all gravitational-wave
observables calculated in the following subsections.

\subsection{Gravitational Wave Phase}
\label{subsec:gw_phase}

Gravitational waves are propagating perturbations of spacetime generated by
time-dependent mass and current multipole moments.  For a nearly circular
EMRI/IMRI the dominant radiation is the quadrupolar $m=2$ harmonic.  Its
instantaneous angular frequency is therefore approximately twice the
orbital angular frequency,
\begin{equation}
 \omega_{\rm GW}(t)=2\Omega[r(t)].
 \label{eq:gw_angular_frequency}
\end{equation}
The accumulated gravitational wave  phase between an initial time $t_i$ and a final time
$t_f$ is
\bqn
 \Phi_{\rm GW}(t_f;t_i)
 &=&\int_{t_i}^{t_f}\omega_{\rm GW}(t)\,dt\nb\\
 &=&2\int_{t_i}^{t_f}\Omega[r(t)]\,dt.
 \label{eq:gw_phase_time}
\eqn
Because the orbital radius decreases monotonically, it is convenient to use
$r$ rather than $t$ as the integration variable.  From
Eq.~\eqref{eq:radial_inspiral}, we find
\begin{equation}
 dt
 =-\frac{dE_{\rm orb}/dr}
 {\mathcal F_{\rm GW}+\mathcal F_{\rm DF}}\,dr.
 \label{eq:dt_dr_phase}
\end{equation}
Thus, for an inspiral from $r_i$ to $r_f<r_i$, we have
\begin{equation}
 \Phi_{\rm GW}(r_i\rightarrow r_f)
 =2\int_{r_f}^{r_i}
 \Omega(r)
 \frac{dE_{\rm orb}/dr}
 {\mathcal F_{\rm GW}+\mathcal F_{\rm DF}}\,dr, 
 \label{eq:gw_phase_radial}
\end{equation}
which is manifestly positive.  The corresponding number of gravitational wave cycles is
$N_{\rm GW}=\Phi_{\rm GW}/(2\pi)$.  Even a small environmental correction to
$\Omega$, $E_{\rm orb}$, or the luminosity may accumulate over the large
number of EMRI/IMRI cycles and produce a measurable dephasing.

To separate the conservative modification of the spacetime from the
dissipative effect of dynamical friction, we calculate three reference
phases over the same observation interval. In all three cases, the inspirals are initialized with the same
orbital radius and gravitational-wave phase and are evolved over
the same observation time. Let
$\Phi_{\rm vac}$ denote the phase in the vacuum Schwarzschild spacetime,
$\Phi_{\rm halo}^{\rm GW}$ the phase in the halo geometry when only gravitational wave
emission is included, and $\Phi_{\rm halo}^{\rm GW+DF}$ the phase in the
same halo geometry when both dissipative channels are included.  We then
define
\bqn
 \Delta\Phi_{\rm geo}
 &\equiv&
 \Phi_{\rm halo}^{\rm GW}-\Phi_{\rm vac},
 \label{eq:phase_geo}
 \\
 \Delta\Phi_{\rm DF}
 &\equiv&
 \Phi_{\rm halo}^{\rm GW+DF}-\Phi_{\rm halo}^{\rm GW},
 \label{eq:phase_df}
 \\
 \Delta\Phi_{\rm tot}
 &\equiv&
 \Phi_{\rm halo}^{\rm GW+DF}-\Phi_{\rm vac}\nb\\
 &=&\Delta\Phi_{\rm geo}+\Delta\Phi_{\rm DF}.
 \label{eq:phase_total}
\eqn  
The first measures
the conservative dephasing caused by the halo-induced modification of the
metric and circular geodesics.  The second isolates the additional
dissipative dephasing caused by DF.  The third is the total environmental
phase shift relative to a vacuum black-hole template.  A negative value
means that the halo model accumulates fewer gravitational wave radians than the reference
model over the chosen observation interval; a positive value means that it
accumulates more. These definitions therefore allow the observable phase shift to be
decomposed unambiguously into conservative and dissipative
environmental contributions.

\subsection{Waveform Mismatch}
\label{subsec:waveform_mismatch}

A phase difference is physically informative, but it does not by itself
state how distinguishable two signals are in a detector.  Detectability is
determined by the full waveform, by the detector noise, and by the freedom
to shift the arbitrary arrival time and initial phase of a template.  The
waveform mismatch provides a standard noise-weighted measure of the loss in
matched-filter signal-to-noise ratio when one waveform is used to represent
another.  In the present context it quantifies how poorly a vacuum waveform
would reproduce a signal generated by an SMBH surrounded by a dark-matter
halo.

Since the orbital evolution is adiabatic, the frequency-domain
waveform can be constructed using the stationary-phase
approximation. For a frequency-domain waveform we write
\begin{equation}
\tilde h(f)=A(f)e^{i\Psi(f)},
\label{eq:fd_waveform}
\end{equation}
where $A(f)$ and $\Psi(f)$ are its amplitude and Fourier phase.  They 
are completely determined by the inspiral trajectory obtained in Sec. III C.

The
noise-weighted inner product between two real detector signals is defined by
\begin{equation}
 (h_1|h_2)
 =4\,\Re\int_{f_{\rm min}}^{f_{\rm max}}
 \frac{\widetilde h_1(f)\widetilde h_2^*(f)}{S_n(f)}\,df,
 \label{eq:noise_inner_product}
\end{equation}
where $S_n(f)$ is the one-sided detector noise power spectral density and an
asterisk denotes complex conjugation.  This weighting suppresses frequency
regions in which the detector is noisy and emphasizes those in which it is
most sensitive.  The norm is $\|h\|=\sqrt{(h|h)}$, and the normalized overlap
is
 \begin{equation}
 \mathcal O(h_1,h_2)
 =\max_{t_c,\phi_c}
 \frac{(h_1|h_2e^{i(2\pi f t_c+\phi_c)})}
 {\sqrt{(h_1|h_1)(h_2|h_2)}}.
 \label{eq:overlap}
\end{equation}
The maximization removes differences due only to an arbitrary time shift
$t_c$ and phase shift $\phi_c$, neither of which carries information about
the dark-matter environment.  
The mismatch is then
\begin{equation}
  {\cal M}(h_1,h_2)=1-\mathcal O(h_1,h_2).
 \label{eq:mismatch}
\end{equation}
Identical waveforms have
$\mathcal O=1$ and $ {\cal M}=0$, whereas a larger mismatch indicates a
larger loss of matched-filter performance.  For a signal with optimal
signal-to-noise ratio $\varrho$, a commonly used distinguishability estimate
is
\begin{equation}
 {\cal M}\gtrsim \frac{1}{2\,\mathrm{SNR}^2},
\label{eq:mismatch_threshold}
\end{equation}
up to details of parameter fitting and the number of intrinsic parameters.
Thus, the mismatch complements the accumulated dephasing: the latter
identifies the physical origin and sign of the phase shift, while the former
assesses its detector-weighted observational significance.

\section{Application to Model I}

Having established the general formalism for circular geodesics, inspiral evolution, gravitational-wave phase accumulation and waveform mismatch in the previous section, we now specialize these results to the exact relativistic dark-matter halo solution (Model I) constructed in \cite{ShenA}. Our goal is to investigate how the spacetime geometry generated by the halo modifies the orbital motion of  EMRIs/IMRIs  and the corresponding gravitational wave signal.

Throughout this section we consider a SMBH with mass
$M=10^6M_\odot$,
whose Schwarzschild radius is
$r_s=2M$.
The characteristic radius of the halo is chosen as
$a=20~{\rm kpc}$,
which is typical for galactic dark-matter halos.

For Model I, the density profile of the dark matter halo is given
by
\begin{equation}
\rho(r)
= \begin{cases}
\frac{\rho_0a^4\left(r- 4M\right)}
{r^2\left(r+a\right)^3},
& r\ge4M, \cr
 0, & r<4M. \cr
 \end{cases}
\end{equation}
The corresponding mass function is
\begin{equation}
m(r)
= M + M_h
\frac{(r-4M)^2}{(r+a)^2}H(r-4M),
\label{eq:ModelI_mass}
\end{equation}
where
\begin{equation}
M_h
\equiv
\frac{2\pi\rho_0a^4}
{a+2r_s}.
\label{eq:Mh_ModelI}
\end{equation}

For fixed $M$ and $a$,  Equation (\ref{eq:Mh_ModelI}) shows that the overall strength of the halo is controlled solely by the characteristic density $\rho_0$.
In the numerical calculations below we consider the representative values
\begin{equation}
\rho_0
=
0.1,\;
0.3,\;
1,\;
3,\;
10
~{\rm GeV\,cm^{-3}},
\end{equation}
which correspond to different values of $M_h$ through Eq.~(\ref{eq:Mh_ModelI}).


\begin{figure}[t]
\centering
\includegraphics[width=0.5\textwidth]{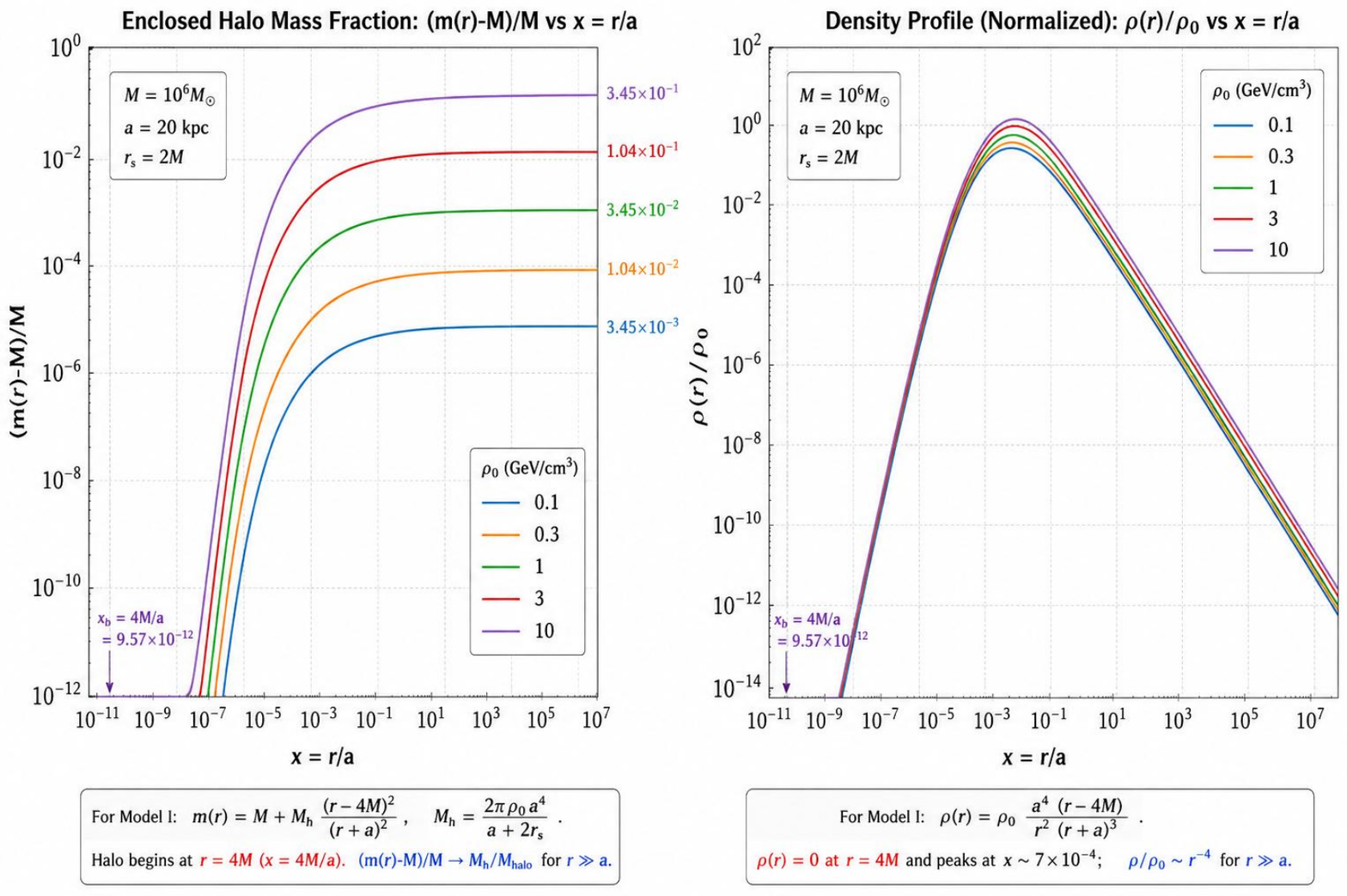} 
\caption{
{\bf Left Panel}: Dimensionless enclosed halo mass,
$(m(r)-M)/M$,
for Model I \cite{ShenA} with different values of the characteristic density $\rho_0$.
The enclosed halo mass vanishes at the inner boundary $r=4M$, increases monotonically outside the horizon, and approaches $M_h/M$ at large distances. Increasing $\rho_0$ changes only the overall amplitude while preserving the functional dependence on the radial coordinate.  {\bf Right Panel}: Energy-density profiles of Model I \cite{ShenA} for the same values of $\rho_0$ as for the mass function.  The density vanishes identically for $r<4M$, rises smoothly outside the horizon, and decreases asymptotically as $r^{-4}$ at large distances.
}
\label{fig1}
\end{figure}

Since the geodesic motion is determined directly by the spacetime geometry, it is useful to first examine the enclosed halo mass,
$[m(r)-M]/M$,
which measures the fractional modification of the Schwarzschild geometry. Figure~\ref{fig1} shows the dimensionless enclosed halo mass for the selected values of $\rho_0$. It is seen that the halo begins at $r=4M$, increases monotonically with the radial coordinate, and asymptotically approaches the total halo mass. Since the orbital frequency, the specific energy, the specific angular momentum and the location of the innermost stable circular orbit all depend directly on the mass function.

 Fig.~\ref{fig1} provides a direct illustration of how the spacetime geometry is modified by the dark-matter halo. It
  presents the corresponding energy-density profiles. Unlike the phenomenological NFW and Hernquist models, the present relativistic solution satisfies the physically desirable requirement that the dark matter vanishes identically inside $r=4M$. Consequently, the spacetime remains exactly Schwarzschild until the halo begins at $r=4M$, avoiding the unphysical accumulation of matter inside the event horizon.

For the EMRIs/IMRIs considered in this work, we restrict attention to the
region
$4M \leq r \lesssim 100\,r_s$.
This interval covers both the relatively early inspiral and the
strong-field portion relevant to gravitational-wave observations.
For the benchmark values $M=10^6M_\odot$ and $a=20~{\rm kpc}$,
one has
\begin{equation}
\frac{r}{a}\lesssim \frac{200M}{a}\approx 5\times10^{-10}\ll1.
\end{equation}
 Therefore, the analytical  expansions of Model I solution developed in Appendix A provide an excellent approximation to the exact spacetime throughout the inspiral. In all the numerical calculations presented below, the orbital evolution is obtained by integrating Eq.~(\ref{eq:radial_inspiral}) using the expanded solution, from which the accumulated phase differences and waveform mismatches are subsequently computed according to the formalism developed in the last section.

 \subsection{Accumulated Gravitational Wave Cycles}
\label{sec:GW_cycles}

We now quantify the accumulated change in the number of gravitational-wave
cycles caused by the conservative modification of the spacetime geometry and
by dynamical friction. For a quasicircular inspiral, the dominant
quadrupolar gravitational-wave frequency is
\begin{equation}
f_{\rm GW}(r)=\frac{\Omega(r)}{\pi},
\end{equation}
where $\Omega(r)$ is the exact circular-orbit frequency derived in
Sec.~\ref{sec:timelike_geodesics}. The accumulated number of
gravitational-wave cycles during an observation interval
$0\leq t\leq T_{\rm obs}$ is therefore
\bqn
N_{\rm GW}(T_{\rm obs})
&=&
\int_0^{T_{\rm obs}} f_{\rm GW}[r(t)]\,dt\nb\\
&=&
\frac{1}{\pi}
\int_0^{T_{\rm obs}}\Omega[r(t)]\,dt .
\label{eq:NGW_time}
\eqn
Using the monotonic inspiral equation,
\begin{equation}
\dot r
= -
\frac{\mathcal F_{\rm GW}+\mathcal F_{\rm DF}}
{dE_{\rm orb}/dr},
\end{equation}
Eq.~\eqref{eq:NGW_time} may equivalently be written as
\begin{equation}
N_{\rm GW}
=
\frac{1}{\pi}
\int_{r_f}^{r_i}
\Omega(r)
\frac{dE_{\rm orb}/dr}
{\mathcal F_{\rm GW}(r)+\mathcal F_{\rm DF}(r)}
\,dr ,
\label{eq:NGW_radial}
\end{equation}
where $r_i$ and $r_f$ are the initial and final orbital radii.

The accumulated number of gravitational-wave cycles is one of
the most sensitive observables for detecting weak environmental
effects, since even a small correction to the orbital evolution
can build up into a measurable phase difference over the long
duration of an EMRI or IMRI inspiral.

To isolate the different environmental effects, we calculate three cycle
counts over the same observation time:
\begin{widetext}
\begin{align}
N_{\rm vac}
&:
\text{Schwarzschild geometry with GW emission only},\\
N_{\rm halo}^{\rm GW}
&:
\text{Model-I geometry with GW emission only},\\
N_{\rm halo}^{\rm GW+DF}
&:
\text{Model-I geometry including both GW emission and DF}.
\end{align}
\end{widetext}
We then define the conservative, dynamical-friction, and total changes in
the number of gravitational-wave cycles as
\bqn
\Delta N_{\rm GW}
&\equiv&
N_{\rm halo}^{\rm GW}-N_{\rm vac},
\label{eq:DeltaN_GW}\\
\Delta N_{\rm DF}
&\equiv&
N_{\rm halo}^{\rm GW+DF}-N_{\rm halo}^{\rm GW},
\label{eq:DeltaN_DF}\\
\Delta N_{\rm tot}
&\equiv&
N_{\rm halo}^{\rm GW+DF}-N_{\rm vac}\nb\\
&=&
\Delta N_{\rm GW}+\Delta N_{\rm DF}.
\label{eq:DeltaN_total}
\eqn
Here $\Delta N_{\rm GW}$ measures the conservative effect of the
dark-matter halo through its modification of the spacetime geometry and
the exact circular geodesics, while $\Delta N_{\rm DF}$ isolates the
additional dissipative contribution from dynamical friction. A negative
value indicates that the corresponding model accumulates fewer
gravitational-wave cycles than the reference model.

For each primary mass, the initial orbital radius is selected so that the
corresponding vacuum inspiral reaches the Schwarzschild ISCO after
$T_{\rm obs}=4\,{\rm yr}$. The same initial radius and observation time are
then used for all Model-I evolutions, allowing a direct comparison of the
accumulated cycles.


\begin{figure}[t]
\centering
\includegraphics[width=0.45\textwidth]{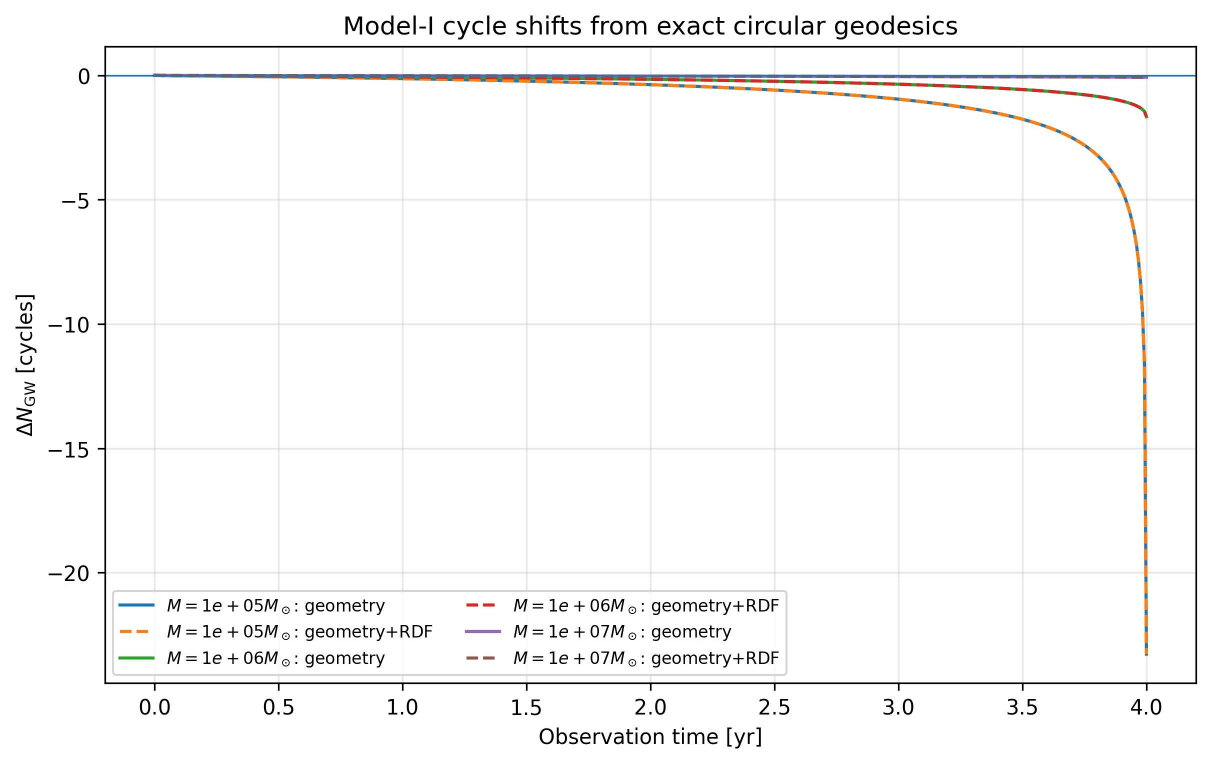}
\caption{Accumulated gravitational-wave cycle shifts over a four-year
observation, computed from the adiabatic inspiral based on the
exact circular geodesics of Model I \cite{ShenA}.   Solid
curves include only the conservative modification of the halo geometry,
whereas dashed curves include both the modified geometry and relativistic
dynamical friction.  Each curve is measured relative to a vacuum
Schwarzschild inspiral with the same masses and initial radius.}
\label{fig2}
\end{figure}

 Figure~\ref{fig2} shows the accumulated changes in the gravitational-wave
cycles over a four-year observation for the representative values
of the characteristic halo density $\rho(r)$. The solid curves
represent the conservative modification of the spacetime geometry,
whereas the dashed curves include both the conservative geometric
effect and the dissipative contribution from relativistic
dynamical friction. Several
features are immediately apparent. First, the accumulated cycle
shift is always negative, indicating that the presence of the halo
reduces the total number of gravitational-wave cycles relative to
the corresponding vacuum inspiral. Second, the magnitude of the
cycle shift increases monotonically with the halo density, since a
denser halo produces a larger modification of the spacetime geometry.
Finally, the solid and dashed curves are almost indistinguishable,
demonstrating that the contribution from relativistic dynamical
friction is negligible compared with the conservative modification
of the background geometry.


\begin{figure}[t]
\centering
\includegraphics[width=0.45\textwidth]{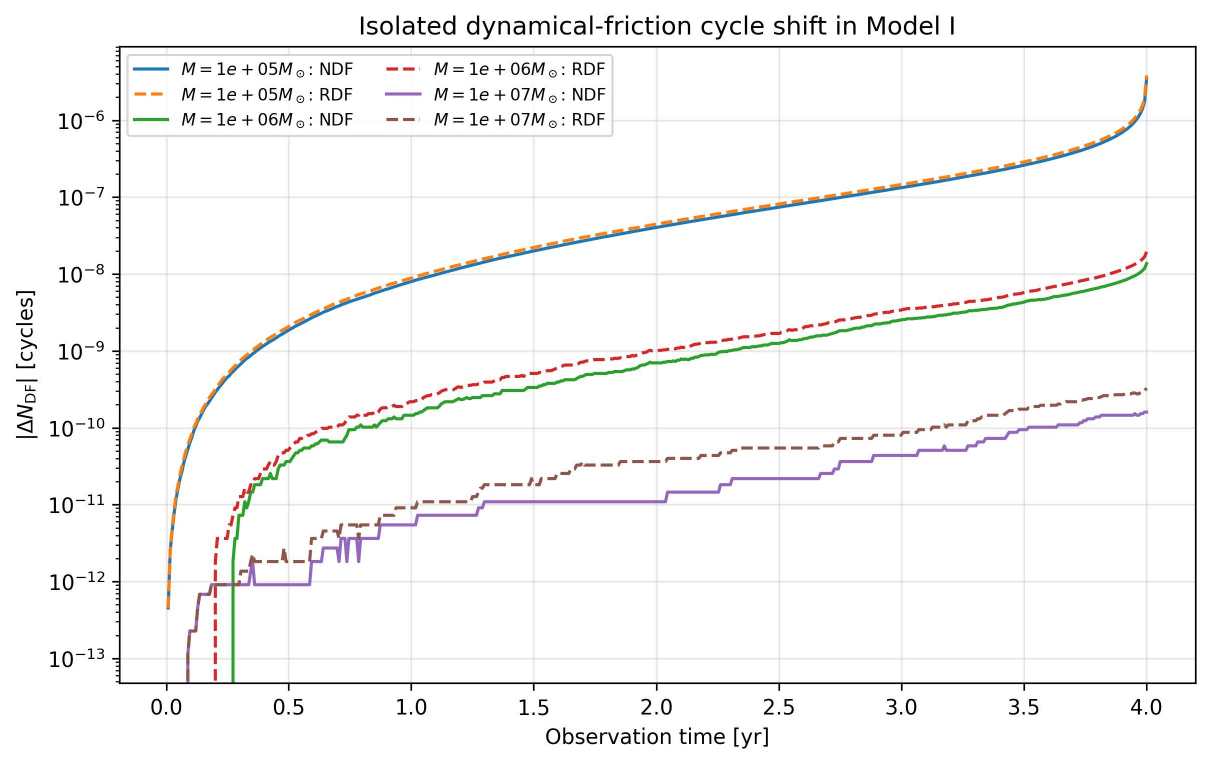}
\caption{Absolute cycle shift produced solely by dynamical friction, obtained
by subtracting the geometry-only Model-I \cite{ShenA} evolution from the evolution with
DF.  Solid and dashed curves correspond to the Newtonian and relativistic DF
prescriptions, respectively.}
\label{fig3}
\end{figure}

Although the dashed and solid curves in
Fig.~\ref{fig2} are nearly indistinguishable,
their small difference can be isolated by considering the
dynamical-friction contribution separately.
Figure~\ref{fig3} therefore shows the
accumulated change in the gravitational-wave cycles produced
solely by relativistic dynamical friction.
 To estimate the observation time required for the dark-matter
environment to produce a potentially measurable gravitational-wave
cycle shift, we define
\begin{equation}
\left|
\Delta N_{\rm tot}\!\left(T_{\rm obs}^{(1)}\right)
\right|=1.
\label{eq:Tobs_one_cycle}
\end{equation}
For each value of the primary mass, the initial radius is chosen so that
the corresponding vacuum inspiral reaches the Schwarzschild ISCO after
an observation time $T_{\rm obs}$. The vacuum and Model-I inspirals are
then evolved from the same initial radius for the same duration, and
Eq.~\eqref{eq:Tobs_one_cycle} is solved numerically.


\begin{figure}[t]
\centering
\includegraphics[width=0.45\textwidth]{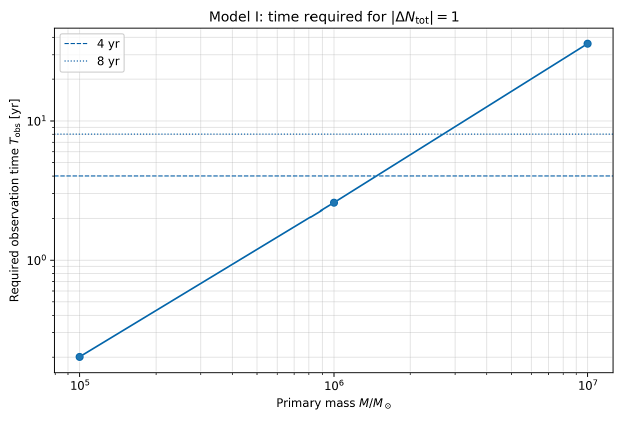}
\caption{Observation time required for the total Model-I \cite{ShenA}  environmental
cycle shift to satisfy $|\Delta N_{\rm tot}|=1$, for a $10M_\odot$
secondary, $a=20\,{\rm kpc}$, $\rho_0=0.3\,{\rm GeV\,cm^{-3}}$, and
$\ln\Lambda=3$. The horizontal lines indicate four- and eight-year
observation periods.}
\label{fig4}
\end{figure}

Figure~\ref{fig4} shows the resulting observation time for
a $10M_\odot$ secondary. The dashed horizontal lines indicate observation
periods of four and eight years. The required observation time increases
rapidly with the primary mass: a one-cycle environmental dephasing is
accumulated after approximately $0.20$ yr, $2.57$ yr, and $35.9$ yr for
$M=10^5M_\odot$, $10^6M_\odot$, and $10^7M_\odot$, respectively.
This behavior reflects the fact that the inspiral proceeds more slowly around more massive SMBHs, so that the environmental modification of the orbital dynamics accumulates more gradually despite the longer orbital timescales.
Consequently, the environmental modification is potentially measurable
within a nominal four-year observation for the two lower primary masses,
whereas a substantially longer observation would be required for
$M=10^7M_\odot$.


\begin{figure}[t]
\centering
\includegraphics[width=0.45\textwidth] {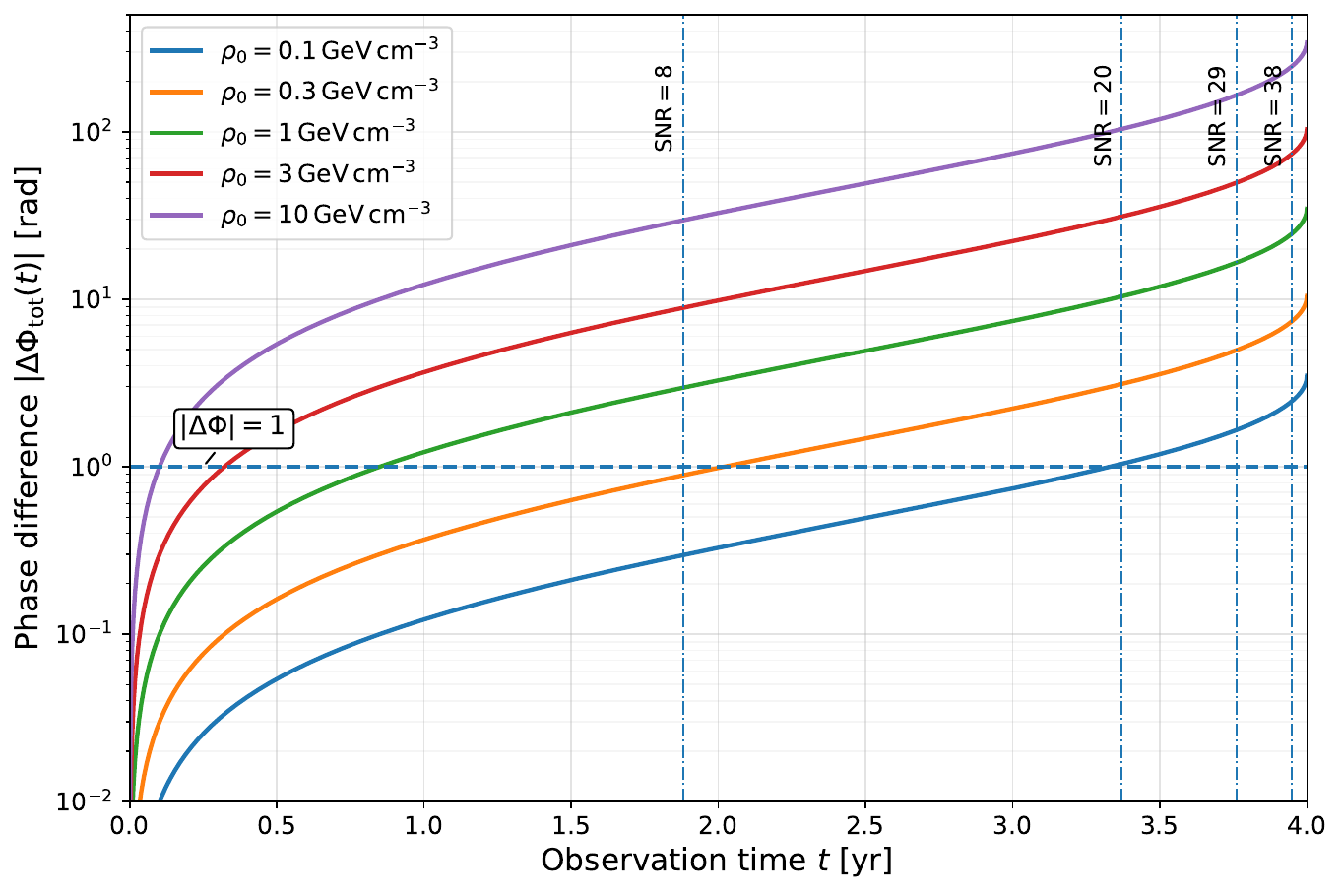}
\caption{
Accumulated gravitational-wave phase difference as a function of
observation time for Model I \cite{ShenA}. The benchmark binary parameters are
$M=10^{6}M_\odot$ and $m_\star=10M_\odot$, while the halo scale radius
is $a=20\,{\rm kpc}$ and the total observation time is
$T_{\rm obs}=4\,{\rm yr}$. The curves correspond to the density
normalizations
$\rho_0=0.1$, $0.3$, $1$, $3$, and
$10\,{\rm GeV\,cm^{-3}}$. In each case,
$\Delta\Phi_{\rm tot}$ is the phase difference between the Model-I
inspiral including relativistic dynamical friction and the corresponding
vacuum Schwarzschild inspiral with the same masses and initial radius.
The horizontal dashed line indicates $|\Delta\Phi|=1$, while the vertical 
dash-dotted lines denote the times at which the accumulated signal-to-noise ratio reaches 
${\rm SNR}(t)=8$, $20$, $29$, and $38$.
}
\label{fig5}
\end{figure}


\begin{figure}[t]
\centering
\includegraphics[width=0.45\textwidth]{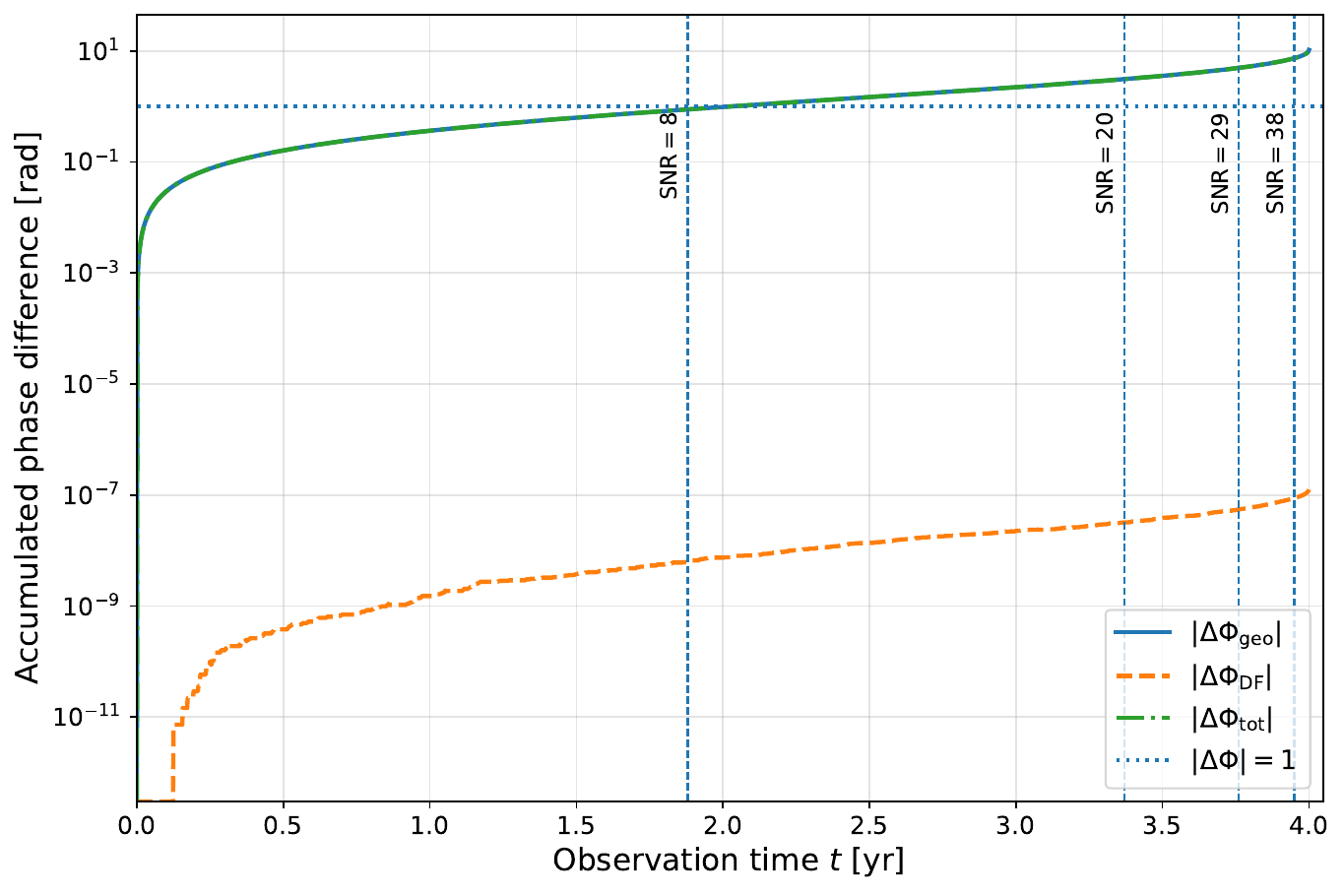}
\caption{
Decomposition of the accumulated gravitational-wave phase difference
for the benchmark Model-I \cite{ShenA}  halo with
$\rho_0=0.3\,{\rm GeV\,cm^{-3}}$.
The three curves show the conservative contribution
$|\Delta\Phi_{\rm geo}|$, the contribution from relativistic
dynamical friction $|\Delta\Phi_{\rm DF}|$, and the total phase
difference $|\Delta\Phi_{\rm tot}|$, respectively.
The horizontal dotted line marks
$|\Delta\Phi|=1$, while the vertical dashed lines indicate the times
at which the accumulated signal-to-noise ratio reaches
${\rm SNR}(t)=8$, $20$, $29$, and $38$.
}
\label{fig6}
\end{figure}

\subsection{Time-Dependent Phase Difference and Accumulated SNR}

The total phase shift accumulated over a complete observation provides
a useful summary of the environmental effect, but it does not show when
during the inspiral the dephasing becomes significant. We therefore
define the time-dependent phase difference
\begin{equation}
 \Delta\Phi_{\rm tot}(t)
 =
 \Phi_{\rm halo}^{\rm GW+DF}(t)-\Phi_{\rm vac}(t),
\end{equation}
where the two signals begin with the same orbital radius and initial
phase. Corresponding conservative and dynamical-friction contributions
are defined by
\begin{align}
 \Delta\Phi_{\rm geo}(t)
 &=
 \Phi_{\rm halo}^{\rm GW}(t)-\Phi_{\rm vac}(t),\\
 \Delta\Phi_{\rm DF}(t)
 &=
 \Phi_{\rm halo}^{\rm GW+DF}(t)
 -
 \Phi_{\rm halo}^{\rm GW}(t).
\end{align}

To indicate which part of the phase evolution contributes most strongly
to the detectable signal, we also calculate the accumulated
signal-to-noise ratio,
\begin{equation}
 {\rm SNR}^2(t)
 =
 4\int_{f_{\rm min}}^{f(t)}
 \frac{|\widetilde h(f')|^2}{S_n(f')}\,df',
\end{equation}
where $S_n(f)$ is the one-sided detector noise power spectral density.
Because $\rho$ is reserved for the dark-matter density, we write the
signal-to-noise ratio explicitly as ${\rm SNR}$ throughout.

While the accumulated change in the number of gravitational-wave
cycles provides an intuitive measure of the environmental effect,
matched-filter searches are sensitive primarily to the accumulated
waveform phase. Even a relatively small change in the orbital
frequency can therefore become observationally significant after
integrating over the millions of gravitational-wave cycles generated
during an EMRI/IMRI inspiral. We now investigate how the accumulated phase difference evolves
throughout the inspiral and when it becomes observationally
significant relative to the accumulated signal-to-noise ratio.

Figure~\ref{fig5} shows the accumulated gravitational-wave phase difference,
$|\Delta\Phi_{\rm tot}(t)|$, as a function of the elapsed observation
time for five representative halo-density normalizations. Several
features are immediately apparent. First, the accumulated phase
difference increases monotonically throughout the inspiral because the
small environmental correction to the orbital frequency continuously
accumulates over the large number of gravitational-wave cycles.
Second, the dephasing depends strongly on the halo density: denser
halos produce a larger modification of the spacetime geometry and
therefore a larger accumulated phase difference. This behavior follows directly from the increase of the enclosed
halo mass with $\rho_0$, which enhances the conservative
modification of the orbital frequency and therefore the cumulative
phase evolution throughout the inspiral.
For the largest density considered here,
$\rho_0=10\,{\rm GeV\,cm^{-3}}$, the accumulated phase
difference exceeds the commonly adopted detectability
threshold of one radian after only a small fraction of
the total observation time, whereas for lower halo
densities the threshold is reached progressively later.

The horizontal dashed line indicates the commonly used
heuristic criterion $|\Delta\Phi|=1$, above which the
environmental dephasing may become potentially observable.
The vertical dash-dotted lines denote the times at which the
accumulated signal-to-noise ratio reaches
${\rm SNR}(t)=8,20,29,$ and $38$, respectively.

The relative ordering of the phase and SNR thresholds depends
strongly on the halo density. For
$\rho_0=1,3,$ and $10~{\rm GeV\,cm^{-3}}$, the accumulated
phase difference exceeds one radian before the signal reaches
${\rm SNR}=8$. For the fiducial density
$\rho_0=0.3~{\rm GeV\,cm^{-3}}$, the one-radian threshold is
reached shortly after ${\rm SNR}=8$, at approximately
$2.02~{\rm yr}$, whereas for
$\rho_0=0.1~{\rm GeV\,cm^{-3}}$ it is reached only after a
substantial fraction of the four-year observation. These
results demonstrate the importance of long-duration
observations for probing lower-density relativistic
dark-matter halos.

To identify the physical origin of the accumulated phase difference,
Fig.~\ref{fig6} decomposes the total phase shift into its conservative
contribution, $\Delta\Phi_{\rm geo}$, arising from the halo-induced
modification of the background spacetime, and the additional
contribution, $\Delta\Phi_{\rm DF}$, produced by relativistic
dynamical friction. Throughout the inspiral,
$|\Delta\Phi_{\rm geo}|$ is essentially indistinguishable from the
total phase difference, whereas
$|\Delta\Phi_{\rm DF}|$ remains many orders of magnitude smaller.
Thus, for the parent-halo profiles considered here, the accumulated
gravitational-wave dephasing is overwhelmingly dominated by the
conservative modification of the spacetime geometry, while the direct contribution from relativistic dynamical friction
remains negligible throughout the inspiral for the halo models
considered in this work.

This conclusion is consistent with the behavior of the accumulated
gravitational-wave cycles discussed in the previous subsection.
Although dynamical friction is included self-consistently in the
orbital evolution, its influence on the observable phase remains far
below that produced by the geometric modification of the Einstein-cloud
spacetime. Consequently, the dominant environmental signature arises from the conservative modification of the background spacetime geometry.

The dominance of the conservative geometric contribution over the dissipative dynamical-friction contribution indicates that the detectability of Einstein-cloud environments is governed primarily by the accumulated waveform dephasing. We therefore turn to the waveform mismatch, which provides a detector-weighted measure of whether these accumulated phase differences can be observationally distinguished.

The accumulated phase differences discussed above provide an
intuitive measure of the environmental imprint on the inspiral.
We next quantify the observational distinguishability of these
signals using the detector-weighted waveform mismatch.

\subsection{Waveform Mismatch}
\label{sec:waveform_mismatch}

The accumulated phase difference discussed in the previous
subsection provides an intuitive measure of the deviation between
two inspiral waveforms. However, their observational
distinguishability is quantified more rigorously by the waveform
mismatch. Following the matched-filtering formalism introduced in
Sec.~III E, we compute the mismatch using
Eqs.~(\ref{eq:noise_inner_product})--(\ref{eq:mismatch}),
maximizing the overlap over a constant relative arrival time
and phase.

Although the accumulated phase difference provides an intuitive
measure of the environmental effect, waveform mismatch offers a
more rigorous detector-weighted criterion for assessing whether
two inspiral signals can be distinguished by matched filtering.

Throughout this work, the gravitational waveform is constructed
directly from the numerical inspiral evolution as
\begin{equation}
h(t)=A(t)\cos\Phi(t),
\label{eq:time_domain_waveform}
\end{equation}
where $A(t)$ and $\Phi(t)$ denote the time-dependent waveform
amplitude and accumulated gravitational-wave phase, respectively.
We retain the full
time dependence of ${\cal A}(t)$ when constructing each waveform. 
Equation~(\ref{eq:time_domain_waveform}) provides the time-domain
representation of the inspiral waveform. The corresponding
frequency-domain waveform used in the matched-filter analysis is
constructed from it using the stationary-phase approximation
introduced in Sec.~III~E.
Although the normalization in Eq.~~(\ref{eq:overlap}) removes the overall
amplitude scale, differences in the relative amplitude evolution
during the inspiral remain encoded in the mismatch through the
noise-weighted inner product. The calculation therefore incorporates both
the phase evolution and the time-dependent amplitude weighting produced by the
orbital dynamics. To separate the different environmental effects, we consider three waveform
mismatches
\begin{align}
{\cal M}_{\rm tot}
&\equiv
{\cal M}
\!\left(
h_{\rm vac},
h_{\rm halo}^{\rm GW+DF}
\right),
\label{eq:mismatch_total}
\\
{\cal M}_{\rm geo}
&\equiv
{\cal M}
\!\left(
h_{\rm vac},
h_{\rm halo}^{\rm GW}
\right),
\label{eq:mismatch_geo}
\\
{\cal M}_{\rm DF}
&\equiv
{\cal M}
\!\left(
h_{\rm halo}^{\rm GW},
h_{\rm halo}^{\rm GW+DF}
\right),
\label{eq:mismatch_df}
\end{align}
where ${\cal M}_{\rm geo}$ measures the conservative
modification of the waveform caused by the
dark-matter halo, ${\cal M}_{\rm DF}$ isolates the
additional contribution from relativistic dynamical
friction, and ${\cal M}_{\rm tot}$ represents the
complete environmental mismatch relative to the
vacuum waveform.
All waveform pairs are generated from inspirals that start from
the same initial orbital radius and initial gravitational-wave
phase and are evolved over the same observation time.

Rather than evaluating the mismatch only at the end of the full four-year
observation, we calculate it as a function of the elapsed observation time. At
each time $T$, the numerical waveforms are restricted to the interval
$0\leq t\leq T$ and multiplied by the same tapering window before their
discrete Fourier transforms are evaluated. Applying an identical window to
both waveforms suppresses spectral leakage associated with the finite
observation interval without introducing a relative modification between the
signals. The resulting time-dependent mismatch is
\begin{equation}
{\cal M}(T)
=
1-
\max_{\Delta t,\Delta\phi}
\frac{
\left(
h_1^{\,T}
\middle|
h_2^{\,T}
\right)
}
{
\sqrt{
\left(
h_1^{\,T}
\middle|
h_1^{\,T}
\right)
\left(
h_2^{\,T}
\middle|
h_2^{\,T}
\right)
}
},
\label{eq:MT}
\end{equation}
where the superscript $T$ denotes a tapered waveform restricted to the
corresponding observation interval. In the numerical implementation, the same
sampling rate, frequency interval, noise curve, and tapering prescription are
used for all waveforms being compared.

For two nearby waveforms with comparable norms, the squared norm of their
difference is approximately related to the mismatch by
\begin{equation}
(\delta h|\delta h)
\simeq
2\,{\rm SNR}^{2}{\cal M},
\label{eq:dist}
\end{equation}
where $\delta h=h_1-h_2$ and ${\rm SNR}$ is the accumulated signal-to-noise
ratio. As a practical fixed-parameter distinguishability criterion, we
therefore use
\begin{equation}
{\cal M}
\gtrsim
\frac{1}{2\,{\rm SNR}^{2}}.
\label{eq:Mcriterion}
\end{equation}
This criterion compares waveforms constructed with the same intrinsic binary
parameters and maximized only over a constant relative time and phase shift. A
full parameter-estimation analysis would additionally allow the intrinsic
source parameters to vary and is beyond the scope of the present work.


\begin{figure*}[t]
\centering
\includegraphics[width=0.45\textwidth]{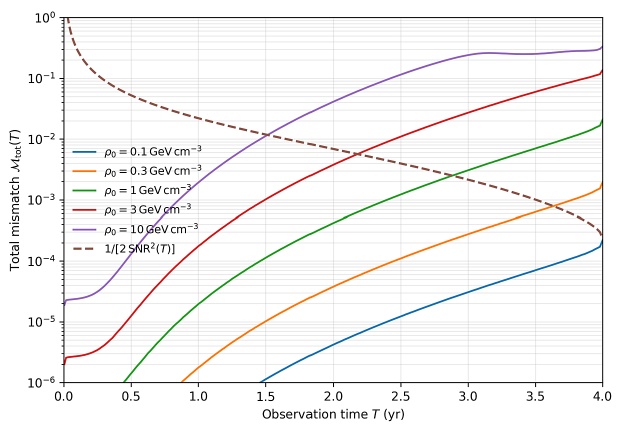}
\caption{
Evolution of the total waveform mismatch,
${\cal M}_{\rm tot}(T)$,
for five dark-matter density normalizations,
$\rho_0=0.1,\;0.3,\;1,\;3$, and
$10~{\rm GeV\,cm^{-3}}$.
The mismatch is computed from the full frequency-domain overlap after maximizing over the relative arrival time and phase.
The dashed curve denotes the distinguishability threshold,
${\cal M}_{\rm th}=1/[2\,{\rm SNR}^{2}(T)]$.
Larger halo densities produce larger waveform mismatches and therefore become
distinguishable after progressively shorter observation times.
}
\label{fig7}
\end{figure*}

Unlike the accumulated phase difference, which depends only on
the phase evolution, the mismatch incorporates both the phase
and amplitude evolution of the waveform through the
noise-weighted detector inner product.

Figure~\ref{fig7} shows the evolution of ${\cal M}_{\rm tot}(T)$ during the four-year
observation for the five benchmark halo-density normalizations,
\begin{equation}
\rho_0
=
0.1,\,
0.3,\,
1,\,
3,\,
10~
{\rm GeV\,cm^{-3}},
\label{eq:mismatch_densities}
\end{equation}
together with the time-dependent distinguishability threshold
\begin{equation}
{\cal M}_{\rm th}(T)
=
\frac{1}{2\,{\rm SNR}^{2}(T)}.
\label{eq:Mthreshold}
\end{equation}
The intersection between a mismatch curve and
${\cal M}_{\rm th}(T)$ determines the earliest observation time at which the
corresponding halo waveform satisfies the adopted fixed-parameter
distinguishability criterion relative to the vacuum waveform.

Figure~\ref{fig7} 
shows that the waveform mismatch
generally increases as the observation proceeds and the
difference between the halo and vacuum waveforms accumulates.
Small nonmonotonic features may arise because the overlap is
recomputed over each finite observation interval and maximized
over the relative arrival time and phase.
The growth rate depends
strongly on the halo density: larger density
normalizations produce larger mismatches and
therefore exceed the distinguishability threshold
after progressively shorter observation times.
 For the largest halo density considered here,
$\rho_0=10~{\rm GeV\,cm^{-3}}$,
the mismatch becomes distinguishable after approximately
$1.5\,{\rm yr}$,
whereas for the fiducial value
$\rho_0=0.3~{\rm GeV\,cm^{-3}}$
the corresponding observation time is about
$3.6\,{\rm yr}$.
By contrast, the weakest halo considered,
$\rho_0=0.1~{\rm GeV\,cm^{-3}}$,
does not reach the distinguishability threshold during the four-year LISA
mission. These results are fully consistent with the accumulated gravitational-wave cycles, phase differences, and accumulated signal-to-noise ratios discussed in the previous subsections,
demonstrating that the waveform
mismatch provides a robust detector-weighted measure of the observational
significance of the dark-matter halo.

Although the total waveform mismatch determines when the dark-matter halo
becomes observationally distinguishable, it does not by itself identify the
physical mechanism responsible for the waveform difference. 
To identify the physical origin of the waveform mismatch, we therefore decompose
 it into its conservative geometric contribution and the additional contribution arising from relativistic dynamical friction.


\begin{figure*}[t]
\centering
\includegraphics[width=0.45\textwidth]{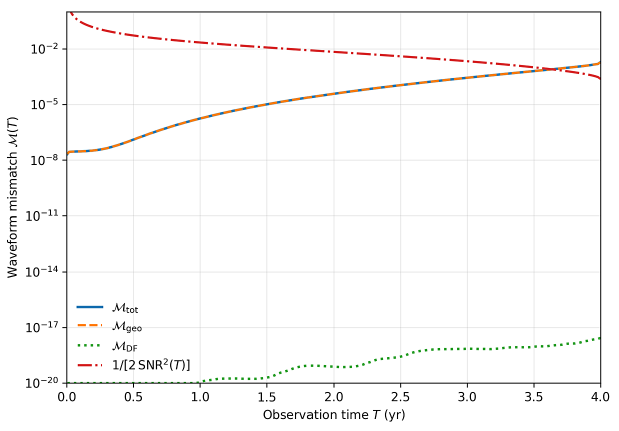}
\caption{
Decomposition of the waveform mismatch for the representative halo density
$\rho_0=0.3~{\rm GeV\,cm^{-3}}$.
The solid curve denotes the total mismatch,
${\cal M}_{\rm tot}$,
while the dashed and dotted curves show the separately defined
geometric and dynamical-friction mismatches,
${\cal M}_{\rm geo}$ and ${\cal M}_{\rm DF}$, respectively.
The dash-dotted curve represents the distinguishability threshold,
${\cal M}_{\rm th}=1/[2\,{\rm SNR}^{2}(T)]$.
The waveform mismatch is dominated by the geometric modification of the
background spacetime, whereas the direct contribution from relativistic
dynamical friction remains negligible throughout the inspiral.
}
\label{fig8}
\end{figure*}

Figure~\ref{fig8} compares the total
mismatch,
${\cal M}_{\rm tot}$,
with its geometric contribution,
${\cal M}_{\rm geo}$,
and the contribution arising solely from dynamical friction,
${\cal M}_{\rm DF}$,
for the representative halo density
$\rho_0=0.3~{\rm GeV\,cm^{-3}}$.
From Fig.~\ref{fig8} it can be seen clearly that the observable waveform mismatch is almost
entirely determined by the conservative modification of the spacetime
geometry. Throughout the inspiral,
${\cal M}_{\rm geo}$
is nearly indistinguishable from the total mismatch,
while the contribution from relativistic dynamical friction remains many
orders of magnitude smaller and therefore has a negligible effect on the
overall detectability of the halo. This result is fully consistent with the
accumulated phase analysis presented above, which likewise showed that the
dominant environmental signature originates from the geometric modification of
the orbital dynamics. Consequently, although relativistic dynamical friction
is included consistently in our calculations, the principal observable effect
of the dark-matter halo is its conservative modification of the background
spacetime.

Taken together, the accumulated gravitational-wave cycles, phase evolution,
signal-to-noise ratio, and waveform mismatch provide mutually consistent
diagnostics of the environmental effects produced by dark-matter halos around
SMBHs. In all cases considered here, the dominant
observable signature originates from the conservative modification of the
spacetime geometry, while the contribution from relativistic dynamical
friction remains comparatively small. These complementary observables
therefore provide a coherent framework for probing dark-matter distributions
around supermassive black holes with future space-based gravitational-wave
detectors.

The waveform-mismatch analysis therefore confirms the
conclusions reached from the accumulated cycle and phase
differences. Across all three diagnostics, the observable
environmental signatures are dominated by the conservative
modification of the spacetime geometry, whereas relativistic
dynamical friction produces only a negligible correction for the
halo models considered here.

\section{Conclusions}

In this paper, we have developed a fully relativistic framework for
studying the gravitational-wave signatures of collisionless
dark-matter halos surrounding SMBHs. Modeling the
halo as a spherically symmetric Einstein cloud, we  
developed a general formalism for investigating their influence on
the orbital evolution and gravitational-wave signatures
of EMRIs/IMRIs embedded in collisionless dark-matter halos.
 As a concrete application, we specialized the general formalism to
Model~I of our previously constructed Einstein-cloud solutions  \cite{ShenA}, and
investigated its effects on the orbital evolution, accumulated
gravitational-wave cycles, waveform dephasing, signal-to-noise ratio,
and waveform mismatch for EMRI/IMRI systems.

The Einstein-cloud model provides a physically motivated description
of collisionless dark matter in the strong-field region surrounding a
SMBH. Unlike many phenomenological halo profiles
that are intended for galactic scales, the present model explicitly
incorporates the relativistic inner cutoff at $r=4M$, as required by
the equilibrium phase-space analyses of collisionless particles in
Schwarzschild spacetime \cite{SFW13,SABB22}. Consequently, it provides a self-consistent
framework for studying relativistic dark-matter environments in the
vicinity of black holes while remaining sufficiently flexible to
describe both cored and cuspy halo distributions through a common
parametrization.

Our analysis demonstrates  that relativistic dark-matter halos can produce
observable imprints on long-duration gravitational-wave signals. We
found that the accumulated gravitational-wave cycles, waveform phase,
and waveform mismatch all increase monotonically throughout the
inspiral and depend sensitively on the halo density. The accumulated
signal-to-noise ratio further determines the observation time at which
these environmental effects become detectable. By decomposing both the
waveform phase and the waveform mismatch into their conservative and
dissipative contributions, we found that the observable signatures are overwhelmingly dominated by the conservative modification of the spacetime geometry, whereas the direct contribution from relativistic dynamical friction remains negligible for the halo models considered here. This conclusion is remarkably robust, being supported independently by
all four observational diagnostics considered in this work: the
accumulated gravitational-wave cycles, waveform phase evolution,
signal-to-noise ratio, and waveform mismatch.
Consequently, the principal observable effect of the
Einstein cloud is encoded in the accumulated gravitational-wave phase.

The results presented here demonstrate that future space-based
gravitational-wave observatories such as LISA provide a promising
opportunity to probe the relativistic distribution of dark matter
around SMBHs. Because the environmental effects
accumulate continuously over the millions of gravitational-wave cycles
generated during an EMRI or IMRI inspiral, even relatively small
departures from the vacuum spacetime may become observationally
significant. Gravitational-wave observations therefore offer a new
window into the strong-gravity distribution of dark matter that is
complementary to electromagnetic observations of black-hole shadows
and stellar dynamics.

Several natural extensions of the present work deserve further
investigation. The formalism developed here can be generalized to
rotating SMBHs and more realistic relativistic
dark-matter halo models. It will also be interesting to investigate
eccentric inspirals, higher-order post-Newtonian corrections,
gravitational self-force effects \cite{ZGW26}, and Bayesian parameter estimation
using realistic detector noise. Such studies will further clarify the
capability of future gravitational-wave observations to constrain the
distribution and physical properties of dark matter in the immediate
vicinity of supermassive black holes.

The framework developed here provides a unified relativistic foundation
for investigating gravitational-wave signatures of Einstein-cloud
environments and establishes gravitational-wave observations as a
promising probe of the strong-field distribution of dark matter around
SMBHs. 

\section*{4acknowledgment}}

F.A. and L.K. thank the Department of Physics $\&$ Astronomy, Baylor University, for the support
through the Department undergraduate summer research program (2026).
A.W.  is partially supported by the
US NSF Grant No. PHY-2308845.

\section*{Appendix A: Expansions of Model I Solutions}
\renewcommand{\theequation}{A.\arabic{equation}}
\setcounter{equation}{0}

We now develop a systematic small-\(x\) expansion of the general-relativistic
Model-I background constructed in \cite{ShenA}, which corresponds to the choice $\left(n, \alpha, \beta, \gamma\right) = (1, 1, 5, 1)$, assuming only
\begin{equation}
x\equiv\frac{r}{a}\ll1,
\end{equation}
without assuming that \(M_h/a\) is small.  Thus \(a\), \(M\), and \(M_h\)
are held fixed when the fields are expanded in powers of \(x\).  It is useful
to introduce
\bqn
\mu&\equiv&\frac{M}{a},\;\;\;
\eta\equiv\frac{M_h}{a},\;\;\; x_b\equiv\frac{4M}{a}=4\mu,\nb\\
\Delta&\equiv& a^2+16MM_h.
\eqn
The physical halo region is \(x\geq x_b\).
For any radial function \(W(r;a,M,M_h)\), we define its dimensionless-variable
representation by
\begin{equation}
 w(x;a,M,M_h)\equiv W(ax;a,M,M_h).
\end{equation}
If necessary, an appropriate power of \(a\) is first extracted so that the
quantity being expanded is dimensionless.  A regular quantity has a Taylor
expansion, while a quantity containing inverse powers of \(r\) generally has a
Laurent expansion,
\bqn
 w(x;a,M,M_h)
 &=& \sum_{n=n_{\min}}^N w_n(a,M,M_h)x^n\nb\\
 && +{\cal{O}}\left(x^{N+1}\right).
\eqn

The exact dimensionless mass function is
\begin{equation}
 \overline m(x)\equiv\frac{m(ax)}{a}
 =\frac{M}{a}+\frac{M_h}{a}\frac{(x-x_b)^2}{(1+x)^2}.
\end{equation}
Using $(1+x)^{-2}=1-2x+3x^2+{\cal{O}}\left(x^3)\right)$, we find
\bqn
 \frac{m(ax)}{a}
 &=&\frac{M}{a}+\frac{M_h}{a}(x-x_b)^2
 \left(1-2x+3x^2\right)\nb\\
 && +{\cal{O}}\left(x^3\right).
\label{eq:m0-boundary-expanded}
\eqn
This form preserves the exact inner-edge condition \(m_0(ax_b)=M\).
Collecting equal powers of \(x\), one may equivalently write
\begin{equation}
 \frac{m_0(ax)}{a}
 =\sum_{n=0}^2 \mathfrak m_n x^n+{\cal{O}}\left(x^3\right),
\end{equation}
where
\bqn
 \mathfrak m_0&=&\frac{M\Delta}{a^3},\;\;\;
 \mathfrak m_1=-\frac{8MM_h(a+4M)}{a^3},\nb\\
 \mathfrak m_2&=&\frac{M_h(a^2+16Ma+48M^2)}{a^3}.
\eqn
The boundary-adapted expression in Eq.~\eqref{eq:m0-boundary-expanded} is
preferable when matching at \(x=x_b\), whereas the collected polynomial is
convenient for coefficient calculations.

On the other hand, the exact density becomes
\begin{equation}
 \kappa a^2\rho(x)
 =4\frac{M_h}{a}\left(1+\frac{4M}{a}\right)
 \frac{x-x_b}{x^2(1+x)^3},
\end{equation}
from which we obtain the
junction-preserving expression
\bqn
 \kappa a^2\rho(x)
 &=&4\frac{M_h}{a}\left(1+\frac{4M}{a}\right)
 \frac{x-x_b}{x^2}\nb\\
 && \times \left(1-3x+6x^2\right) +{\cal{O}}\left(x^3\right). ~~~~~~~~
\label{eq:rho0-boundary-expanded}
\eqn 
It makes \(\rho(x_b)=0\) manifest.  In Laurent form,
\begin{equation}
 \kappa a^2\rho(x)
 =\sum_{n=-2}^{2}\varrho_n x^n +{\cal{O}}\left(x^3\right),
\end{equation}
with
\bqn
 \varrho_{-2}&=&-\frac{16MM_h(a+4M)}{a^3},\nb\\
 \varrho_{-1}&=&\frac{4M_h(a+4M)(a+12M)}{a^3},\nb\\
 \varrho_0&=&-\frac{12M_h(a+4M)(a+8M)}{a^3},\nb\\
 \varrho_1&=&\frac{8M_h(a+4M)(3a+20M)}{a^3},\nb\\
 \varrho_2&=&-\frac{40M_h(a+4M)(a+6M)}{a^3}.
\eqn
Although the Laurent series is algebraically useful, the factored form
Eq.~\eqref{eq:rho0-boundary-expanded} should be used when the exact zero at the
inner edge is important.

The exact pressure relation can be written in either of the equivalent forms
\begin{equation}
 P=\frac{\rho m}{2(r-2m)}
 =\frac{\rho}{4}\left(\frac{r}{r-2m}-1\right).
\label{eq:P0-useful-identity}
\end{equation}
Define
\begin{equation}
 g(x)\equiv x-\frac{2m(ax)}{a}.
\end{equation}
Then
\begin{equation}
 \kappa a^2P(x)
 =\frac{\kappa a^2\rho(x)}{4}
 \left[\frac{x}{g(x)}-1\right].
\label{eq:P0-boundary-form}
\end{equation}
Using Eq.~\eqref{eq:m0-boundary-expanded}, we find
\bqn 
 g(x)&=& x
 -2\frac{M_h}{a}(x-x_b)^2(1-2x+3x^2)\nb\\
 && -\frac{2M}{a} +{\cal{O}}\left(x^3\right).
\eqn
At the boundary, \(g(x_b)=2M/a\neq0\); hence
Eq.~\eqref{eq:P0-boundary-form} and \(\rho(x_b)=0\) imply
\(P(x_b)=0\).

The corresponding Laurent expansion is
\begin{equation}
 \kappa a^2P(x)
 =\sum_{n=-2}^{2}p_nx^n +{\cal{O}}\left(x^3\right).
\label{eq:P0-Laurent}
\end{equation}
The coefficients are
\begin{widetext}
\bqn
 p_{-2}&=&\frac{4MM_h(a+4M)}{a^3},\;\;\;
 p_{-1}=
 \frac{M_h(a+4M)\left[a^2(a-12M)-16MM_h(a+12M)\right]}
 {a^3\Delta},\nb\\
 p_0&=&
 \frac{M_h(a+4M)}{2Ma^3\Delta}
 \left[a^2(a^2-6Ma+48M^2)+96M^2M_h(a+8M)\right],\nb\\
 p_1&=&
 \frac{M_h(a+4M)}{4M^2a^3\Delta^2}
 \Bigl\{
 a^4(a^3-6Ma^2+24M^2a-160M^3)
 \nonumber\\
 &&
 +8MM_ha^2(a^3-12Ma^2-640M^3)
 -2048M^4M_h^2(3a+20M)
 \Bigr\},
\nb\\
 p_2&=&
 \frac{M_h(a+4M)}{8M^3a^3\Delta^2}
 \Bigl\{
 a^4(a^4-6Ma^3+24M^2a^2-80M^3a+480M^4)
 \nonumber\\
 &&
 +4MM_ha^2(a^4-12Ma^3+96M^2a^2+3840M^4)
 +20480M^5M_h^2(a+6M)
 \Bigr\}.
 \label{eq:ptwo}
\eqn
\end{widetext}
The factored expression \eqref{eq:P0-boundary-form}, rather than the truncated
Laurent polynomial by itself, should be used to enforce the inner-edge zero.

The field equation for the temporal metric function is
\begin{equation}
 \frac{d\ln f}{dx}
 =\frac{2\overline m(x)}{x[x-2\overline m(x)]}
 =-\frac1x+\frac{1}{g(x)}.
\label{eq:f0-x-equation}
\end{equation}
Using the collected mass coefficients above, the local solution is
\begin{equation}
 f(x)=\frac{C_+}{x}
 \left(1+\alpha_1x+\alpha_2x^2\right) +{\cal{O}}\left(x^3\right),  
\label{eq:f0-exterior-expanded}
\end{equation}
where
\begin{equation}
 \alpha_1=-\frac{a^3}{2M\Delta},
 \;\;\;
 \alpha_2=-\frac{2a^3M_h(a+4M)}{M\Delta^2}.
\label{eq:alpha12}
\end{equation}
The constant \(C_+\) is the exterior normalization.  In the complete exact
solution it is fixed by the chosen normalization at large radius, normally
asymptotic flatness.  The local small-\(x\) expansion alone does not determine
\(C_+\).

For \(r<4M\), the matter fields vanish and the mass function is constant,
\begin{equation}
 m(r)=M,\qquad \rho(r)=P(r)=0.
\end{equation}
The interior temporal metric function contains an independent constant,
\begin{equation}
 f(r)=C_{\rm in}\left(1-\frac{2M}{r}\right).
\label{eq:f-interior}
\end{equation}
Because the field equations are first order in \(m(r)\) and \(f(r)\), we impose
continuity of the fields themselves at the inner edge,
\begin{equation}
 [m]_{4M}=[\rho]_{4M}=[P]_{4M}=[f]_{4M}=0,
\label{eq:junction-fields}
\end{equation}
without introducing continuity of their first derivatives as an independent
condition.  Equations \eqref{eq:m0-boundary-expanded},
\eqref{eq:rho0-boundary-expanded}, and \eqref{eq:P0-boundary-form} give
\begin{equation}
 m(4M)=M,\quad \rho(4M)=0,\quad P(4M)=0.
\end{equation}
Continuity of \(f\) gives
 \begin{equation}
 C_{\rm in}=2f_{0+}(x_b).
\label{eq:Cin-general}
\end{equation}
Applying the exterior expansion \eqref{eq:f0-exterior-expanded},
\begin{equation}
 C_{\rm in}
 =\frac{2C_+}{x_b}
 \left(1+\alpha_1x_b+\alpha_2x_b^2\right)+{\cal{O}}\left(x_b^2\right).
\label{eq:Cin-expanded}
\end{equation}
Equivalently, after substituting \(x_b=4M/a\),
\bqn
 C_{\rm in}
 &=&\frac{aC_+}{2M}
 \left[1-\frac{2a^2}{\Delta}
 -\frac{32aMM_h(a+4M)}{\Delta^2}\right]\nb\\
 && + {\cal{O}}\left(x_b^2\right). 
\label{eq:Cin-explicit}
\eqn

\end{document}